%% file: main.tex
\documentclass[acmsmall, nonacm]{acmart} 
\usepackage{enumitem}
\usepackage{amsfonts}
\usepackage{mathrsfs} 
\usepackage{amsmath} 
\usepackage{bm}
\usepackage{mathtools}
\usepackage{mathpartir} 
\usepackage{bbding}
\usepackage{color}
\usepackage[ruled,noend]{algorithm2e}

\SetCommentSty{algComments}

\usepackage[capitalize]{cleveref}
\usepackage{xparse} 
\usepackage{ifthen} 
\usepackage{thmtools} 
\usepackage{thm-restate} 
\usepackage{bm}
\usepackage{subcaption}
\usepackage{wrapfig}
\usepackage{tikz} 
\usepackage{newfile} 
\usepackage{totcount} 
\usepackage{array} 
\usepackage{subcaption}
\usepackage{multirow}
\usepackage{nicematrix} 

\regtotcounter{@todonotes@numberoftodonotes}

\usepackage{listings}

  \lstdefinestyle{tinyc}{
    basicstyle=\scriptsize\ttfamily,
    keywordstyle=\color{blue}
  }

  \lstdefinestyle{normalc}{
    basicstyle=\ttfamily,
    numbers=none,
    keywordstyle=\color{blue}
  }

  \lstdefinestyle{inlinec}{
    basicstyle=\ttfamily
  }

  \lstdefinelanguage{Golang}%
  {morekeywords=[1]{package,import,func,type,struct,return,defer,panic,%
     recover,select,var,const,iota,},%
   morekeywords=[2]{string,uint,uint8,uint16,uint32,uint64,int,int8,int16,%
     int32,int64,bool,float32,float64,complex64,complex128,byte,rune,uintptr,%
     error,interface},%
   morekeywords=[3]{map,slice,make,new,nil,len,cap,copy,close,true,false,%
     delete,append,real,imag,complex,chan,},%
   morekeywords=[4]{for,break,continue,range,go,goto,switch,case,fallthrough,if,%
     else,default,},%
   morekeywords=[5]{Println,Printf,Error,Print,},%
   sensitive=true,%
   morecomment=[l]{//},%
   morecomment=[s]{/*}{*/},%
   morestring=[b]',%
   morestring=[b]",%
   morestring=[s]{`}{`},%
   }

\usetikzlibrary{backgrounds}
\tikzstyle{every picture}+=[remember picture]

\usepackage{todonotes}

\usepackage{tikz}
\usetikzlibrary{automata}
\tikzset{gadget/.style={->,>=stealth,initial text=,minimum size=7pt,auto,on grid,scale=1,inner sep=1pt,node distance=1cm}}
\tikzset{every state/.style={minimum size=15pt,inner sep=1pt,fill=black!10,draw=black!70,thick}}

\usetikzlibrary{backgrounds,calc}

 \usetikzlibrary{ shapes, fit, arrows}
 \usetikzlibrary{decorations} 
\DeclareMathSymbol{\mdot}{\mathord}{symbols}{"01}
\usepackage{lscape}

\usepackage[utf8]{inputenc}
\usepackage[T1]{fontenc}
\usepackage{microtype}
\usepackage{bussproofs}
\usepackage{tikzit}
\input{tikzit_styles.tikzstyles}

\input{macros}

\title{Reward Augmentation in Reinforcement Learning for Testing Distributed Systems}

\newcommand{\MPIInstitution}{Max Planck Institute for Software Systems (MPI-SWS)}
\newcommand{\MPIStreet}{Paul-Ehrlich-Stra{\ss}e, Building G26}
\newcommand{\MPICity}{Kaiserslautern}
\newcommand{\MPIPostcode}{67663}
\newcommand{\MPICountry}{Germany}

\author{Andrea Borgarelli}
\orcid{0009-0002-4842-6599}             
\affiliation{
  \institution{\MPIInstitution}            
  \streetaddress{\MPIStreet}
  \city{\MPICity}
  \postcode{\MPIPostcode}
  \country{\MPICountry}                    
}
\email{aborgare@mpi-sws.org}

\author{Constantin Enea}
\affiliation{
\institution{CNRS, LIX, Ecole Polytechnique}
\country{France}
}
\email{cenea@lix.polytechnique.fr}

\author{Rupak Majumdar}
\orcid{0000-0003-2136-0542}             
\affiliation{
  \institution{\MPIInstitution}            
  \streetaddress{\MPIStreet}
  \city{\MPICity}
  \postcode{\MPIPostcode}
  \country{\MPICountry}                    
}
\email{rupak@mpi-sws.org}          

\author{Srinidhi Nagendra}
\orcid{}             
\affiliation{
  \institution{Chennai Mathematical Institute, India, IRIF, CNRS, Universite Paris Cite}
  \country{France}           
}

\input{abstract}



\begin{document}

\makeatletter
\newtheoremstyle{dazzle}%
{.5\baselineskip\@plus.2\baselineskip
  \@minus.2\baselineskip}
{.5\baselineskip\@plus.2\baselineskip
  \@minus.2\baselineskip}
{\@acmplainbodyfont}
{\@acmplainindent}
{\bfseries}
{.}
{.5em}
{\thmname{\textcolor{red}{\textbf{#1}}}\thmnumber{ \textcolor{red}{\textbf{#2}}}\thmnote{ {\@acmplainnotefont(\textcolor{blue}{#3})}}}
\makeatother

\theoremstyle{dazzle}
\newtheorem{maintheorem}[theorem]{Theorem}
\newtheorem{mainlemma}[theorem]{Lemma}
\newtheorem{maincorollary}[theorem]{Corollary}
\newtheorem{mainproposition}[theorem]{Proposition}

\crefalias{maintheorem}{theorem}
\crefalias{mainlemma}{lemma}
\crefalias{maincorollary}{corollary}
\crefalias{mainproposition}{proposition}

\theoremstyle{acmplain}
\newtheorem{observation}[theorem]{Observation}
\theoremstyle{acmdefinition}
\newtheorem{remark}[theorem]{Remark}

\Crefname{observation}{Observation}{Observations}
\newcolumntype{H}{>{\setbox0=\hbox\bgroup}c<{\egroup}@{}}

\renewcommand{\lstlistingname}{\listingsInLatex}

\maketitle

\input{contents/intro}

\input{contents/overview}
\input{contents/transition_system}
\input{contents/hierarchies}
\input{contents/eval}
\input{contents/related_work}

\input{contents/conclusion}
\sloppy 

\newpage
\section*{Data-Availability Statement}

The software implementing the two algorithms \pureExpAlgo{} and \biasExpAlgo{} that we use to present our evaluation \cref{sec:eval} is available on Github\footnote{https://www.github.com/zeu5/raft-rl-test}\footnote{https://www.github.com/zeu5/dist-rl-testing} and on Zenodo~\footnote{https://zenodo.org/doi/10.5281/zenodo.12671211}~\cite{zenodo-artifact}.

\section*{Acknowledgements}

We thank Burcu Kulahcioglu Ozkan, Abhik Roychoudhury, and the anonymous reviewers of OOPSLA 2024 for reviewing and providing constructive feedback for this paper. 
Constantin Enea was partially supported by ANR award SCEPROOF and
Rupak Majumdar was supported in part by the the Deutsche Forschungsgemeinschaft project 389792660 TRR 248--CPEC.

\label{beforebibliography}
\newoutputstream{pages}
\openoutputfile{main.pages.ctr}{pages}
\addtostream{pages}{\getpagerefnumber{beforebibliography}}
\closeoutputstream{pages}
\bibliography{bibliography}

\label{afterbibliography}
\newoutputstream{pagesbib}
\openoutputfile{main.pagesbib.ctr}{pagesbib}
\addtostream{pagesbib}{\getpagerefnumber{afterbibliography}}
\closeoutputstream{pagesbib}

\newoutputstream{todos}
\openoutputfile{main.todos.ctr}{todos}
\addtostream{todos}{\arabic{@todonotes@numberoftodonotes}}
\closeoutputstream{todos}

\label{endofdocument}
\newoutputstream{pagestotal}
\openoutputfile{main.pagestotal.ctr}{pagestotal}
\addtostream{pagestotal}{\getpagerefnumber{endofdocument}}
\closeoutputstream{pagestotal}


\end{document}

%% file: tikzit_styles.tikzstyles
\tikzstyle{plain circle}=[fill=white, draw=black, shape=circle]
\tikzstyle{black circle}=[fill=black, draw=black, shape=circle]
\tikzstyle{blue circle}=[fill=blue, draw=black, shape=circle]
\tikzstyle{red circle}=[fill=red, draw=black, shape=circle]
\tikzstyle{blue-green circle}=[fill={rgb,255: red,0; green,128; blue,128}, draw=black, shape=circle]

\tikzstyle{direction edge}=[->]
\tikzstyle{dashed directional edge}=[->, dashed]
\tikzstyle{red full edge}=[draw=red, ->]
\tikzstyle{dashed simple edge}=[-, dashed]
\tikzstyle{double directional}=[double, ->]
\tikzstyle{dash double}=[dashed, double, ->]
\tikzstyle{new edge style 0}=[-, dashed, draw=red]
\tikzstyle{blue}=[-, draw=blue]
\tikzstyle{blue directiona}=[draw=blue, ->]
\tikzstyle{blue dashed}=[dashed, draw=blue, ->]

%% file: macros.tex
\DeclareDocumentCommand{\autstep}{O{}}{%
        \xrightarrow{#1}%
        }
\DeclareDocumentCommand{\langof}{O{} m}{%
  \mathsf{L}_{#1}(#2)%
  }

\newcommand{\listingsInLatex}{Code}

\newcommand{\node}{\mathsf{node}}

\newcommand{\stateSpace}{\mathcal{S}}
\newcommand{\actionSpace}{\mathcal{A}}
\newcommand{\colors}{\mathcal{C}}

\newcommand{\pureExpAlgo}{\textsc{BonusMaxRL}}
\newcommand{\biasExpAlgo}{\textsc{WaypointRL}}
\newcommand{\negRL}{\textsc{NegRL}}
\newcommand{\negRLVisits}{\textsc{NegRLVisits}}

\def\Final#1{{#1}}

%% file: abstract.tex
\begin{abstract}
Bugs in popular distributed protocol implementations have been the source of many downtimes in popular internet services. 
We describe a randomized testing approach for distributed protocol implementations based on reinforcement learning.
Since the natural reward structure is very sparse, the key to successful exploration in reinforcement learning is reward augmentation.
We show two different techniques that build on one another.
First, we provide a decaying exploration bonus based on the discovery of new states---the reward decays as the same state is visited multiple times.
The exploration bonus captures the intuition from coverage-guided fuzzing of prioritizing new coverage points; in contrast to other schemes, we show that
taking the maximum of the bonus and the Q-value leads to more effective exploration.
Second, we provide waypoints to the algorithm as a sequence of predicates that capture interesting semantic scenarios.
Waypoints exploit designer insight about the protocol
and guide the exploration to ``interesting'' parts of the state space.
Our reward structure ensures that new episodes can reliably get to deep interesting states even without execution caching.
We have implemented our algorithm in Go.
Our evaluation on three large benchmarks (RedisRaft, Etcd, and RSL) shows that our algorithm can significantly outperform baseline approaches
in terms of coverage and bug finding.
\end{abstract}

%% file: contents/intro.tex
\section{Introduction}
\label{sec:intro}

Distributed protocols form the core components in many large-scale software systems today.
These protocols enable new, innovative, and business-critical applications.
The appications rely on correct implementations of the protocols
to ensure fundamental correctness properties such as data consistency and durability even in the presence of machine and network faults of various kinds.
Correct and performant implementation of complex protocols, such as those for consensus or distributed transactions, is very difficult, since the programmer
must account for the many possible failure and recovery scenarios.
Indeed, there is a rich literature that demonstrates the presence of many bugs in even well-tested and production-ready implementations.

Many different bug-finding techniques have been applied to distributed systems implementations.
Systematic exploration techniques based on model checking~\cite{DBLP:conf/nsdi/YangCWXLLYLZZ09} attempt to explore the entire space of concurrency
and faults; however, for industrial implementations, model checkers only cover a very small fraction of behaviors.
In practice, the exploration partial, and guided by \emph{search policies} that prioritize certain behaviors over others.
The simplest search policy is pure random exploration \cite{jepsen,DBLP:conf/asplos/BurckhardtKMN10,DBLP:conf/cav/YuanYG18,DBLP:journals/pacmpl/OzkanMO19}:
set up the system and perform random operations, and inject network failures or crashes at random.
Randomized exploration is surprisingly effective in finding bugs, but the search is ``blind.'' 
It is often improved by providing coverage-based feedback (fuzzing) \cite{DBLP:conf/icse/GaoDWFWZH23,DBLP:conf/pldi/Sen08,DBLP:conf/icse/WangSG11}.
Fuzzing techniques based on structural coverage (e.g., line or branch coverage) are very successful in
many domains of sequential software, but their effectiveness in the distributed setting is less understood.
Finding good search policies is an important challenge.

In this paper, we cast the problem of finding optimal search policies in distributed systems as a \emph{reinforcement learning} (RL) problem and propose new algorithms for biased exploration
of the program state space.

In reinforcement learning (see, e.g., \cite{DBLP:books/lib/SuttonB98}), an agent interacts with an environment by looking at the state of the environment and picking a possible action.
The action causes an update to the state through a possibly probabilistic environment transition, and the agent receives a reward based on the state and the action.
The agent interacts with the environment in a series of episodes, where each episode is a fixed finite-length run of the system.
The goal of the agent is to learn an \emph{optimal policy} (a strategy to pick actions based on state-feedback) that maximizes its expected utility,
where the utility is defined as some cumulation of the rewards obtained along the way.
RL techniques have achieved significant successes in many domains in recent years, and it is natural to study how and whether they improve testing.

There is a straightforward correspondence of RL concepts to testing: the agent is the exploration engine and the environment is the system under test.
The state of the system is a partial observation of the system under test---such as contents of logs, messages, or the control state of each process.
An action is the delivery of one or more messages, a process or network crash, or recovery. 
The goal of the agent is to learn a policy to effectively schedule network events or failures to drive the system to a bug.
So far, so good. 
The difficulty in the correspondence is the sparsity of rewards.
If the agent gets a reward only upon a bug, then most episodes will yield no reward (and moreover, one can stop as soon as a single reward is found).
Since policy learning in RL propagates rewards backward in time, the algorithm will revert to random exploration if it does not obtain rewards on most episodes.

RL with sparse rewards has been studied both in the theoretical RL community and in testing.
General reward-free exploration using RL is well studied 
with strong theoretical optimality results~\cite{DBLP:conf/icml/JinKSY20,DBLP:conf/nips/ZhangMS20}. 
These algorithms \emph{augment} the original, possibly sparse, reward with a time-decaying reward that encourages exploration.
However, the theoretical guarantees do not directly translate to optimal performance in practice for a number of reasons. 
First, the number of episodes required to achieve the optimal policy is polynomial in the size of the state and action spaces and the length of the horizon.
When applied to testing distributed systems, the required number of episodes is too large.
Second, the exploration strategy with reward augmentation is similar to a depth-first search approach, as the algorithm attempts to learn the transition relation.
In testing practice, depth first search provides low coverage within a fixed budget of episodes and performs poorly as a bug finding strategy.

In the context of testing, a common approach is to augment the state space with punishments to bias the search to broad exploration
\cite{DBLP:journals/pacmpl/MukherjeeDBL20,DBLP:conf/ccs/MengPRS23,DBLP:conf/icse/ReddyLPS20}.
For example, \citet{DBLP:journals/pacmpl/MukherjeeDBL20} propose an ``always punish'' strategy: in each step, no matter what action is chosen, the agent is punished by getting a negative
reward.
Similarly, \citet{DBLP:conf/icse/ReddyLPS20,DBLP:conf/ccs/MengPRS23} assign a zero reward whenever a coverage goal is met for the first time (a ``bonus''), and a negative reward otherwise. 
The negative rewards force the learning algorithm to explore other alternatives, leading to broad exploration and higher coverage.
Not providing a punishment when a new coverage goal is met for the first time prioritizes search around that point, akin to coverage-guided fuzzing.
These approaches are strongly biased to exploration.
\Final{
However, they do not use any semantic information to guide the search.
}

\subsubsection*{\textbf{\emph{Our Contribution}}}

Our work builds on prior approaches but improves the state of the art in reward augmentation in two ways.
First, we use insights from the theoretical results in reward-free reinforcement learning \cite{DBLP:conf/icml/JinKSY20,DBLP:conf/nips/ZhangMS20} 
to propose a reward augmentation scheme that\Final{, in practice,} works surprisingly well in exploring the state
space (\pureExpAlgo{}). 
Like reward-free schemes, we provide a time-decaying bonus: 
when a coverage goal is reached for the first time, the agent receives a bonus of 1, but the bonus decreases as $1/k$ as the goal is reached $k$ times.
This prioritizes new parts of the state space in the short run, but the decay forces the agent to explore other actions if the local search around a newly 
uncovered goal does not lead to further new goals. 
However, unlike reward-free approaches, we propagate the bonuses using the \textit{max} function instead of adding them, 
ignoring bonus values along the path leading to a new state. 
This focuses the search to target new states over balancing exploration. 
\Final{Therefore, \pureExpAlgo{} behaves differently from prior theoretical results and is tailored for maximizing exploration of unique states within a fixed episode budget.}

Second, we augment the exploration using \Final{programmer-provided} \emph{waypoints}---semantically interesting program states that guide the search to interesting scenarios---which are 
used to provide additional rewards to the agent (\biasExpAlgo{}).
The waypoints are interesting protocol states that one can get from the description of the protocol; they encode developer insight about interesting ``progress points'' in the protocol.
For example, in a consensus protocol, a waypoint might state that a leader has been elected or that one or more entries have been committed.
We show that an RL agent can be biased towards interesting protocol states by providing a sequence of waypoints.
In fact, even partial hints that state generic properties of a protocol, but are agnostic to the actual implementation details, are already effective in guiding the search.
Moreover, when combined with the coverage bonus of \pureExpAlgo{}, we show that intermediate paths between waypoints are maintained during the learning process, so future episodes can 
traverse the initial portions of the path to the final waypoint in a directed way.
In contrast, if we only provided waypoints, a random search or a punishment strategy does not exploit the waypoints in future episodes.

Empirically, we show that the combination of coverage-based augmentation and waypoints empirically outperforms other approaches in terms of coverage and bug finding. 
We implement and test three implementations: RedisRaft, Etcd, and RSL. 
We successfully bias exploration towards 20 out of 26 different waypoints. 
As a result, we uncover a total of 13 bugs (5 new) in the 3 benchmarks with higher frequency than other approaches in a statistically significant way. 
\Final{Since waypoints allow a developer to bias the search to specific parts of the state space,
biased exploration (even when no bugs are found) increases developer confidence 
in the correctness of specific parts of the codebase.}

\Final{To apply \pureExpAlgo{} and \biasExpAlgo{}, we model a distributed system as a Markov Decision Process (MDP). 
In the process, we explore different models while optimizing for coverage. 
We define a model that is parametric in the state abstraction and the granularity of actions.
We pick a concrete model that strikes a balance between allowing RL algorithms to learn while exploring meaningful states. 
We list generic guidelines for controlling the parameters to achieve high coverage. 
The resulting model and general guidelines we provide are an additional contribution of our work.}

\Final{In what follows, we first describe (Section~\ref{sec:overview}) \pureExpAlgo{} and \biasExpAlgo{} with a generic introduction to reinforcement learning. 
Then, we provide a gridworld analogy (Section~\ref{sec:cubes}) explaining the working of the two algorithms. The gridworld also provides an intuition of the state space of distributed systems.}

%% file: contents/overview.tex
\section{Reinforcement Learning with Coverage Bonus and Waypoints}
\label{sec:overview}

Given a distributed system implementation, our goal is to explore its state space to find bugs.
We guide the exploration by controlling the network (the order in which messages get delivered) and by introducing failures (crashes or network partitions) or by 
recovering from previous failures.
An exploration algorithm learns a \emph{policy} (which action to take at a given point) over time, by performing repeated explorations of finite-length executions, seeing the results,
and adapting its choice of actions based on prior results.
We formalize the exploration algorithm as an agent and the learning of a policy as a reinforcement learning task.
We first provide background on reinforcement learning and then describe our algorithms \pureExpAlgo{} and \biasExpAlgo{}.

\subsection{Background: Reinforcement Learning and Q-Learning}
\label{sec:sys_model}

In reinforcement learning (see, e.g., \cite{DBLP:books/lib/SuttonB98}), an agent explores an environment modeled as a Markov decision process.
The agent starts from a fixed initial state---in our setting, an initial configuration of processes---and performs a series of actions---in our setting, message deliveries, faults, or recoveries.
An action updates the state of the system and also provides a reward.
The goal of the agent is to learn a policy that maximizes its expected reward.

Formally, the environment is modelled as a Markov Decision Process (MDP). An MDP consists of 
    a state space $\stateSpace$,
    an action space $\actionSpace$,
    an initial state $s_0 \in \stateSpace$,
    a transition probability function $\mathcal{T}(s,a,s'): \stateSpace \times \actionSpace \times \stateSpace  \rightarrow [0,1]$;
    we write $\mathcal{T}(s,a,s')$ to denote the probability of the transition $s \xrightarrow{a} s'$,
    and a reward function $\mathcal{R}: \stateSpace \times \actionSpace \times \stateSpace \rightarrow \mathbb{R}$ that specifies a reward for a specific transition.
Our goal is to learn a policy $\pi: \stateSpace \rightarrow \Delta(\actionSpace)$ that maps each state to a probability distribution over the actions $\actionSpace$ so that
the expected discounted sum of rewards is maximized.

\input{contents/generic_rl.tex}

Algorithm~\ref{alg:generic_rl} describes a generic Reinforcement Learning loop.
It takes as input the number of iterations (called \emph{episodes}) $K$, the length of each execution (called \emph{horizon}) $H$,
the environment (MDP) being explored $E$, and an RL agent $A$.

The behavior of the environment is captured by the following funcations:
$\mathit{reset}$ resets the MDP to its initial state,
$\mathit{actions}$ returns the subset of actions available at a state,
and $\mathit{step}$ performs one transition from the current state $s$ and the chosen action $a$
according to the (unknown) transition function $\mathcal{T}$ and returns a new state $s'$ by sampling the distribution $\mathcal{T}(s, a,\cdot)$ and a reward that depends on
the state and the action.
Each iteration starts from the environment's initial state, i.e., $\mathit{state}^k_1 = s_0$ for all $k$.

At each step, the RL agent selects the next action from the set of possible actions, according to its policy, and observes the resulting state and reward.
It tracks the current trace and updates its policy based on the rewards collected in the trace.
\Final{The agent is characterized by the following functions:
the function $\mathit{newEpisode}$, called at the beginning of each episode initializes its data structures for the current episode, 
the function $\mathit{recordStep}$, called after each timestep updates the current trace with the current step, 
and $\mathit{processEpisode}$, called at the end of each episode updates the current policy based on the current trace. 
Different instantiations of these functions lead to different RL agents.
} 

The goal of the RL agent is to maximize the expected reward collected throughout the episode, by learning an optimal policy. 
To learn such a policy, the agent iteratively updates its own policy according to the observed transitions and rewards.

A popular RL algorithm is $Q$-Learning~\cite{DBLP:journals/ml/WatkinsD92}. In $Q$-Learning, the agent keeps an estimate $Q(s,a)$ of the expected reward for a 
given state ($s$) and action ($a$) pair called its $Q$-value. 
For a given transition $(s,a,s')$, and its associated reward $r(s,a,s')$, 
$Q$-learning updates the Q-value as follows:
\[Q(s,a) = (1-\alpha) \cdot Q(s,a) + \alpha \cdot (r(s,a,s') + \gamma \cdot \max\limits_{a'} Q(s',a'))\]
where $\alpha, \gamma \in [0,1]$ are hyper-parameters representing the learning rate and discount factor respectively. 
Intuitively, each time the outcome of picking a state-action pair is observed, its $Q$-value is updated based on the observed reward and estimated 
value of the next state. 
\Final{The hyper-parameter} $\alpha$ determines how quickly the $Q$-value changes at each update, and $\gamma$ defines how quickly delayed rewards decrease in value.

The collection of all the $Q$-Values (one for each state-action pair) is called the $Q$-Table. 
The agent's policy at a state $s$ picks an action in proportion to the values in $Q(s,\cdot)$.
Specifically, we use an $\epsilon$-greedy strategy: 
the agent picks the action with the highest $Q$-value at $s$ with probability $1-\epsilon$ 
and picks a random action with probability $\epsilon$,
where $\epsilon$ is a hyper-parameter of the algorithm.

For testing distributed systems, the main technical difficulty is to define the reward function. As explained earlier, assigning rewards only to bad states is too sparse.
We define rewards in two steps: we provide an \emph{exploration bonus} to the agent if they discover a new state (Algorithm~\pureExpAlgo{})
and we use \emph{waypoints} to guide the search (Algorithm~\biasExpAlgo{}).
We describe these algorithms next.

In both the algorithms, we update the data structures at the end of an episode to backpropagate the updated values faster and achieve higher efficiency in exploration. 
Assuming states are, in general, not repeated throughout an episode, updating at each step would require several episodes to back propagate an updated value to the initial state. 
Updating backwards at the end of the episode, instead, allows for an updated reward to be back propagated all the way in a single sweep, thanks to the order of the updates.

\subsection{\pureExpAlgo}

\input{contents/bonus_max.tex}

Algorithm~\ref{alg:bonus_max} shows the implementation of the \pureExpAlgo{} exploration policy. 
The hyper-parameters $\alpha$, $\gamma$, and $\epsilon$ are given as input. The $pick$ function, which is the way the policy chooses the next action, is the standard $\epsilon$-greedy function. 
\Final{Among the state's available actions, with probability $\epsilon$, the policy will return a random action and with probability $1-\epsilon$ the policy will return the action corresponding to the highest $Q$-value}

The $\mathit{processEpisode}$ function shows how the policy is updated to maximize the novelty of observed states. 
Note that no reward is coming from the environment and updates to Q-Values are entirely based on the internal exploration bonus of the policy. 
The policy keeps a value of visits $V(s,a)$ for each observed state-action pair, and provides a reward that is inversely proportional to this value. 
\Final{\pureExpAlgo{} contains two major differences to existing approaches.}

\Final{
    First, the exploration reward is inversely proportional to the number of visits. 
Specifically the reward is $\frac{1}{t}$ where $t$ is the number of visits. 
The visits are recorded by the policy in a table $V(s,a)$ and for every transition $(s,a,s'$), the visits are updated by incrementing $V(s,a)$. 
This reward mechanism rewards new states (with a reward of $1$) and diminishes the reward as the number of visits increases.
Second, the update rule of \pureExpAlgo{} is:
}
\[Q(s,a) = (1-\alpha) \cdot Q(s,a) + \alpha \cdot max(r, \gamma \cdot \max\limits_{a'} Q(s',a'))\]
\Final{Note the use of $max$ instead of the traditional addition. With this update rule, the Q-Value will now be an estimation based only on the best (least visited) reachable state from that state-action pair. In practice, our algorithm will prioritize a path leading to a new state while ignoring how many times the other states along the path have been visited. When no new states are reachable along a path, the value of its states will converge towards 0 as the number of visits increase.}

Our update rule, based on \textit{maximum} instead of sum of immediate and future rewards, does not lead to an optimal policy in a traditional RL setting, 
where an explicit reward function is provided. 
\pureExpAlgo{} could possibly ignore smaller reward signals along trajectories and therefore be unable to learn an optimal policy. 
For example, consider an environment with a reward function such that two paths lead to the same reward signal. 
However one of the paths is longer and contains a small additional reward along the way. 
\pureExpAlgo{} would learn to follow the sub-optimal shorter path even though the longer path leads to higher total reward. 
\Final{Despite the shortcoming, \pureExpAlgo{} performs better when testing distributed systems. 
The reason is that the greedy approach of prioritizing new immediate states aligns well with the goal of pure exploration.}

\subsection{\biasExpAlgo}

The \biasExpAlgo{} algorithm allows to specify target states in the form of \textit{predicates} that they should satisfy, thus allowing the search to be guided by semantic knowledge about states. 
\biasExpAlgo{} takes a sequence of predicates $\{pred_1, \cdots, pred_n \}$ as input. The last predicate $pred_n$ defines the target space to explore, $pred_1$ is the starting predicate, always true, while the others can be used to guide the exploration towards that target space.

The core idea is to maintain a separate exploration $Q$-table for each predicate and use them according to which predicates are satisfied. In addition to the exploration bonus, $Q$-values are updated with rewards whenever the agent progresses towards the target predicate in the predicates list. In other words, using the basic exploration algorithm, the agent will learn a policy to reach the target state space and then maximize its exploration.  

\input{contents/predh_functions.tex}

The specific implementation of \biasExpAlgo{} is shown in Algorithm~\ref{alg:bias_exp_algo} and ~\ref{alg:bias_exp_algo_2}. At each timestep, the highest indexed predicate, among the ones that are true in the current state, is the active predicate $p$. \biasExpAlgo{} will pick an action using the $Q$-table corresponding to $p$ and store the transition $(s,a,s',p,p')$ in the episode trace, where $p$ and $p'$ are the active predicates respectively in $s$ and $s'$. In the $processEpisode$ method, the agent will update the $Q$-tables using the episode trace. First, it checks if the trace reached the final predicate and eventually stores the step where it happened. Then, it will go backwards in the trace transitions and, for each tuple $(s,a,s',p,p')$, it will update the corresponding $Q$-value $Q_p(s,a)$. If the predicate did not change ($p = p'$), the update is equivalent to the \pureExpAlgo{} update, using only the exploration bonus for values updates. On the other hand, if the predicate changed in the transition, the update can involve additional rewards. Specifically, a reward for progressing $progR$ is given if the agent reached an active predicate $p'$ with higher index, meaning that it got closer to the target space. Additionally, if it reached the target space later on in the episode, another bonus $finalR$ is given, discounted according to the number of steps required to reach the target space. Put differently, every time the state space corresponding to a predicate is left, that predicate's policy is updated with positive rewards if it led to higher indexed predicate and if it subsequently reached the target space in the episode. The role of $progR$ is to incentivize immediate progress towards next predicates, while the role of $finalR$ is to improve overall optimization towards the target space. We fixed a value of 2 for these rewards and 1 for the maximum exploration bonus, but these values can be treated as hyper-parameters to control how much the agent will explore before converging to a fixed policy to reach the target state. 
\input{contents/predh_functions_2.tex}

As input, together with the hyper-parameters $\alpha, \gamma$ and $\epsilon$, a boolean $oneTime$ is given. If $oneTime$ is true, once $pred_n$ has been satisfied, all subsequent states in the episode are considered part of the target space. 
This is used to specify a target space related to an occurring event, without having to include the entire history in the state representation. For example, we might be interested in exploring states after visiting a state with specific features. By setting $oneTime$ to true, reaching the specified state will set the $reached$ flag to true and, within the episode, subsequently visited states will be considered part of the target space and be explored using the the $pred_n$ corresponding $Q$-table.

A different $Q$-table for each predicate is initialized in the $init$ method. The $newEpisode$ method resets the trace and the \textit{reached} flag, and checks the active predicate for the initial state. $Pred_1$ is the constant $\top$ predicate. The $pick$ method follows the $\epsilon$-$greedy$ approach on the $Q$-table of the active predicate. In the $recordStep$ method, the agent updates the active predicate for the next state, eventually setting the \textit{reached} flag to true if it reached the target predicate, and it appends the transition $(s,a,s',p,p')$ to the trace. 

\subsection{Intuition: Exploring a Cube world}
\label{sec:cubes}

To demonstrate the efficacy of our algorithms, let us consider a ``cube world'' consisting of a set of 3 dimensional cubes that we want to explore. Each cube in the set is subdivided into a three-dimensional cube of fixed dimension with width ($w$), breadth ($b$), and depth ($d$). The state space is thus a 4-tuple $\stateSpace = (g,w,b, d)$ where $g$ defines the cube number in the set. An agent exploring the cubes starts at $(0,0,0,0)$, and at each cell can 
pick one of the following actions:  \textit{up}, \textit{down}, \textit{left}, \textit{right}, \textit{above}, \textit{below}, \textit{into}, and \textit{reset\_depth}. 
The directions will result in moving by one cell, \textit{into} allows to transition through a door, if present, and \textit{reset\_depth} brings back the agent to depth zero. 
The cubes are connected by special cells that act as doors. At doors, the agent can move uni-directionally \textit{into} the next cube by picking the corresponding action. 
In our example, with 6 $10\times10\times6$ cubes, we place doors such that $(0,5,5,d) \xrightarrow{into} (1,0,0,0)$, $(1,5,5,d) \xrightarrow{into} (2,0,0,0)$, and so on, for any depth $d$ of the cubes. 
The structure and actions of this example aim to reproduce dynamics that can resemble the ones of a system's execution, such as irreversible transitions (doors) or state resets (\textit{reset\_depth} action).

\begin{figure}[t]
    \centering
    \begin{subfigure}{\textwidth}
        \centering
        \includegraphics[width=\linewidth]{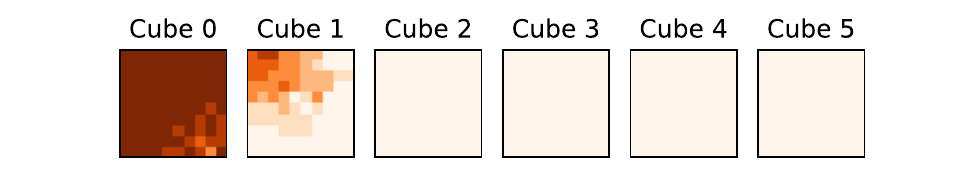}
        \caption{Random exploration}
    \end{subfigure}%
    \vskip\baselineskip
    \begin{subfigure}{\textwidth}
        \centering
        \includegraphics[width=\linewidth]{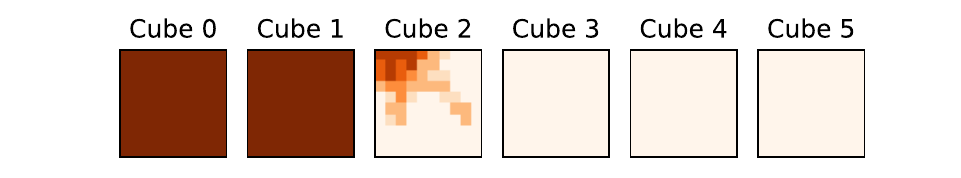}
        \caption{\pureExpAlgo{} exploration}
        \label{subfig:grid_exploration_bonusMax}
    \end{subfigure}%
    \vskip\baselineskip
    \begin{subfigure}{\textwidth}
        \centering
        \includegraphics[width=\linewidth]{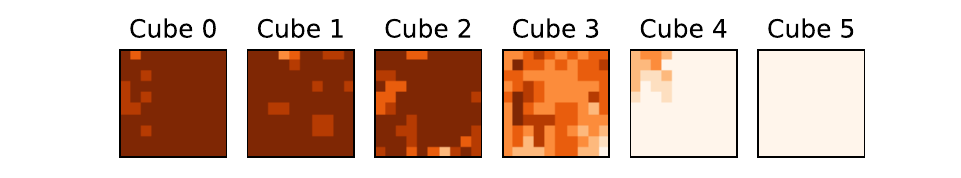}
        \caption{\pureExpAlgo{} exploration with depth abstraction}
        \label{subfig:grid_exploration_bonusMaxAbs}
    \end{subfigure}%
    \vskip\baselineskip
    \begin{subfigure}{\textwidth}
        \centering
        \includegraphics[width=\linewidth]{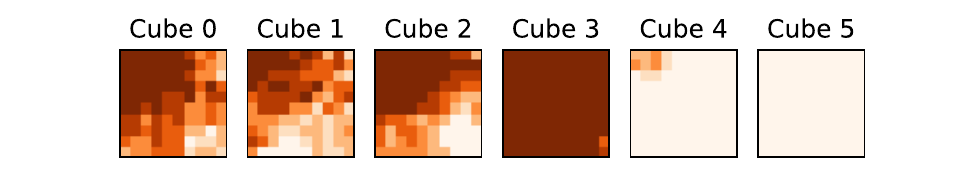}
        \caption{\biasExpAlgo{} exploration \Final{(target cube 3)}}
        \label{subfig:grid_exploration_WaypointRL}
    \end{subfigure}%

    \caption{Exploration of a $6\times10\times10\times6$ cube world, with a given episode budget, using different agents. We plot the heatmap of the top of each cube. The intensity is the sum of \Final{the visited cells along the depth of the cube, with the darkest color meaning all the cells have been visited. Here we showcase several points. First, \pureExpAlgo{} (b) achieves better exploration than Random (a), covering more cells. Second, unbiased exploration struggles to reach cubes away from the starting point (b). Third, chosing an appropriate state space abstraction can lead to better coverage, but it can result in reduced capabilities of systematically exploring a target subspace (c), while \biasExpAlgo{} is able to effectively bias the exploration towards the target cube and almost fully cover it (d).}}
    \label{fig:grid_exploration}
\end{figure}

We evaluate different exploration agents to explore the cubes and illustrate \Final{the outcome} in Figure~\ref{fig:grid_exploration}. From the starting state, an agent takes a fixed number of steps (horizon) and repeats this process for a \Final{given} number of episodes. As clear from the Figure, \pureExpAlgo{} covers significantly more cells than \Final{the} random exploration agent.\footnote{
We run the exploration with a horizon of 80 and for 5k episodes. The hyper parameters for \pureExpAlgo{} are $\alpha = 0.3, \gamma = 0.99$} 
\Final{Additionally}, the results show that an RL agent can fail to completely cover cubes that are farther from the initial state.

\begin{figure}[t]
    \centering
    \begin{subfigure}{\textwidth}
        \centering
        \includegraphics[width=\linewidth]{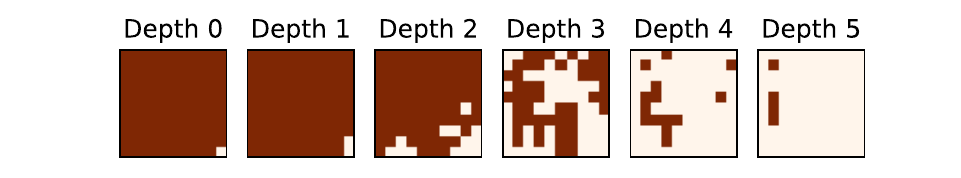}
        \caption{\Final{\pureExpAlgo{} exploration with depth abstraction}}
        \label{fig:grid_cube3_exploration_abs}
    \end{subfigure}%
    \vskip\baselineskip
    \begin{subfigure}{\textwidth}
        \centering
        \includegraphics[width=\linewidth]{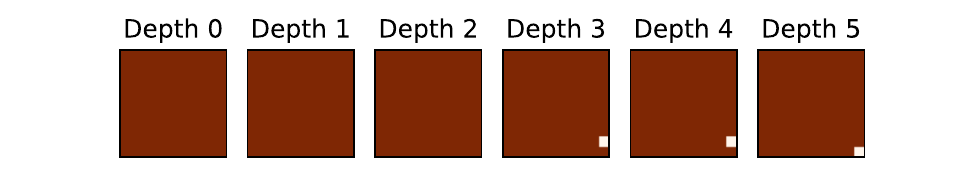}
        \caption{\biasExpAlgo{} cube 3 exploration}
        \label{fig:grid_cube3_exploration_WP}
    \end{subfigure}%
    \caption{\Final{Detailed Exploration of cube 3 in the $6\times10\times10\times6$ cube world. Each grid represents a depth level of the cube. The colored cells have been explored by the agent. (a) \pureExpAlgo{} using the depth abstraction (b) \biasExpAlgo{} with reaching cube 1, 2, and 3 as waypoints. \biasExpAlgo{} is able to explore almost all the cells of the cube.}}
    \label{fig:grid_exploration_cube3}
\end{figure}

Let us suppose we are interested only in covering all cells of a specific cube \Final{(}e.g. cube 3\Final{)}. A trivial solution would be to bias exploration towards cube 3 with an additional \Final{bonus} reward \Final{when reaching the target cube. As shown in Figure~\ref{subfig:grid_exploration_bonusMax}, the reward would not help, since the agent is never able to reach cube 3 within the given episode budget and hence it would never collect the additional reward.} Alternatively, we can improve coverage of cube 3 by introducing an abstraction on the state space. With the abstraction, the state space that RL should cover is smaller and therefore we expect \Final{a better coverage of cube 3}. An example abstraction would be \Final{to ignore the depth co-ordinate, since it is irrelevant to navigate through the cubes (the doors are located along the whole depth of the cubes).} Now, RL will explore more cubes but will \Final{not systematically} explore all depths of each cube. In other words, we lose the granularity of the step. \Final{Figure~\ref{subfig:grid_exploration_bonusMaxAbs} shows the cube world coverage of \pureExpAlgo{} using the defined abstraction. It is able to reach and explore part of cube 3, but without covering its entire depth.}

\Final{Within the same budget,} our solution \biasExpAlgo{} achieves the best coverage in cube 3 (Figure~\ref{subfig:grid_exploration_WaypointRL}). In \biasExpAlgo{}, we split the task into two. First, reaching \Final{the target space (cube 3)} and second, exploring once the target is reached.  We use a different $Q$ table for each task and provide independent rewards. \biasExpAlgo{} uses \Final{the corresponding $Q$ table to pick} actions until cube 3 is reached and subsequently \Final{performs pure exploration.} \Final{The first task can be further split into subtasks by providing additional waypoints to guide the agent.} \biasExpAlgo{} generalizes the reward strategy and accepts a set of target state predicates as waypoints. The algorithm associates each waypoint with a specific $Q$-table that is used to pick actions and update rewards. \Final{In our example, we provided reaching cube 1 and 2 as additional waypoints to guide the agent.}  With \biasExpAlgo{}, we are able to \Final{reach the target space faster}, while retaining the granularity of the exploration step. \Final{Figure~\ref{subfig:grid_exploration_WaypointRL} shows how \biasExpAlgo{} avoids exploring the previous cubes and achieves efficient exploration of the target subspace, while Figure~\ref{fig:grid_exploration_cube3} shows in detail the depths coverage of cube 3 for \pureExpAlgo{} with abstraction and \biasExpAlgo{}.} \footnote{We run the exploration with a horizon of 80 and for 5k episodes. The hyper parameters for \biasExpAlgo{} are $\alpha = 0.3, \gamma = 0.99$}

As clear from the cube example, RL based agents effectively explore states and also are able to bias exploration. The state space of a distributed program is analogous to the cubes. Distributed protocols have communicating processes that reach different states based on different message inter-leavings. Additionally, when there is a quorum of messages or when a timeout occurs, the processes make irreversible progress (akin to moving through a door). In Section~\ref{sec:eval} we will replicate similar results for 3 different distributed protocol implementations. 

%% file: contents/generic_rl.tex
\begin{algorithm}[t]
    \SetAlgoLined
    \SetKwInOut{Input}{Input}
    \Input{$K$: number of episodes, $H$: episode horizon, $E$: environment, $A$: agent}
    \BlankLine
    
    $A.init()$ \\
    \BlankLine

    \For{episode $k=1, \cdots, K$}{
        $state_1^k \gets E.reset()$ \\
        $A.newEpisode(state_1^k) $ \\
        \For{step $h=1, \cdots, H$}{
            $action_h^k \gets A.pick(state_h^k, E.actions(state_h^k))$ \\
            $state_{h+1}^k, reward_h^k \gets E.step(action_h^k)$ \\
            $A.recordStep(state_h^k, action_h^k, state_{h+1}^k, reward_h^k)$ \\
        }
        $A.processEpisode()$ \\
    }
    \caption{Generic RL loop}
    \label{alg:generic_rl}
\end{algorithm}

%% file: contents/bonus_max.tex
\begin{algorithm}[t]
    \SetAlgoLined
    \DontPrintSemicolon
    \SetKwInOut{Input}{Input}
    \SetKwProg{Def}{def}{:}{}
    \Input{$\alpha, \gamma, \epsilon$}
    \BlankLine

    \Def(\tcp*[f]{initialize the Q-Table}){$init()$}{
		$Q(s,a) \leftarrow 1, V(s,a) \leftarrow 0$ for all $s \in \stateSpace, a \in \actionSpace$\;
	}
	\BlankLine
	 \Def(\tcp*[f]{reset the trace}){$newEpisode(\_)$}{
		$trace \gets []$\;
	}
	\BlankLine
    \Def(\tcp*[f]{$\epsilon-greedy$ choice of action}){$pick(s,actions)$}{
    	$x \sim U(0, 1)$ // sample a value x uniformly at random (u.a.r.) from (0,1) \;
        \If {$x < \epsilon$} {
            \Return{$ a \sim U\{actions\} $} // return an element from $actions$ chosen u.a.r. \;
        }
        \Else{
            \Return{$arg max_{a} Q(s, a)$}\;
        }
    }
    \BlankLine
     \Def{$recordStep(state, action, newState, \_ )$}{
    	$trace \gets append(trace, (state, action, newState))$\;
    }
    \BlankLine
    \Def(\tcp*[f]{backward traversal to update Q-Values}){$processEpisode()$}{
        \For{$ i = length(trace) \cdots 1 $}{
        	$(s,a,s') \gets trace[i] $ \;
            $t \gets V(s,a)+1$\;
            $V(s,a) \gets t$\;
            $r \gets \frac{1}{t}$\;
            \If{$ i < length(trace) $}{
            	$Q(s,a) \gets (1-\alpha) \cdot Q(s,a) + \alpha \cdot max(r, \gamma \cdot \max_{a'}Q(s',a'))$ \;
            }
            \Else{
            	$Q(s,a) \gets (1-\alpha) \cdot Q(s,a) + \alpha \cdot max(r, 0)$ \;
            }
        }
    }
	\BlankLine
    \caption{\pureExpAlgo{}: Positive reward based exploration algorithm}
    \label{alg:bonus_max}
\end{algorithm}

%% file: contents/predh_functions.tex
\begin{algorithm}[t]
    \SetAlgoLined
    \DontPrintSemicolon
    \SetKwInOut{Input}{Input}
    \SetKwProg{Def}{def}{:}{}
    \Input{ $ predicates = \left\{pred_1, \cdots, pred_n\right\}, oneTime \in \left\{\top, \bot\right\}, \alpha, \gamma, \epsilon$}
    \BlankLine

    \Def{$init()$}{
    	\For(\tcp*[f]{init a Q-Table for each predicate}){$i = 1 \cdots n$}{
    		$Q_i(s,a) \leftarrow 1, V_i(s,a) \leftarrow 0$ for all $s \in \stateSpace, a \in \actionSpace$\;
    	}
	}
	\BlankLine
	\Def(\tcp*[f]{reset values and active predicate}){$newEpisode(initialState)$}{
		$trace \gets []$,
		$reached \gets \bot $ \;
		\For{$i = n \cdots 1$}{
			\If{$predicate_i(initialState) = \top $}{
				$activePredicate \gets i$ \;
				$break$ \;
			}
		}
	}
	\BlankLine
	\Def{$pick(s,actions)$}{
		$x \sim U(0, 1)$ \;
		\If {$x < \epsilon$} {
			\Return{random $a \sim U(actions)$}\;
		}
		\Else(\tcp*[f]{pick greedy w.r.t. current predicate Q-Table}){
			$ i \gets activePredicate$ \;
			\Return{$arg max_{a} \: Q_{i}(s, a)$}\;
		}
	}
	\BlankLine
	\Def{$recordStep(s, a, s', \_)$}{
		\If(\tcp*[f]{check the new active predicate}){$reached = \bot$}{
			\For{$i = length(predicates) \cdots 1$}{
				\If{$predicate_i(s') = \top $}{
					$nextActivePredicate \gets i$ \;
					$break$ \;
				}
			}
			\If{$nextActivePredicate = n \land oneTime = \top $}{
				$ reached \gets \top $ \;
			}
		}
		\Else(\tcp*[f]{one-time and reached, target predicate active for the rest of the episode}){
			$ nextActivePredicate \gets n $ \;
		}
		
		$trace \gets append(trace, (s, a, s',activePredicate, nextActivePredicate))$\;
		$activePredicate \gets nextActivePredicate $ \;
	}
	\BlankLine
    \caption{\biasExpAlgo{} - $init, newEpisode, pick$, and $recordStep$ methods}
    \label{alg:bias_exp_algo}
\end{algorithm}

%% file: contents/predh_functions_2.tex
\begin{algorithm}[t]
	\SetAlgoLined
	\DontPrintSemicolon
	\SetKwInOut{Input}{Input}
	\SetKwProg{Def}{def}{:}{}
	\Input{ $ predicates = \left\{pred_1, \cdots, pred_n\right\}, oneTime \in \left\{\top, \bot\right\}, \alpha, \gamma, \epsilon$}
	\BlankLine
	
	\Def{$processEpisode()$}{
		\For(\tcp*[f]{check if the episode reached the target predicate}){$ i = 1 \cdots length(trace)$}{ 
			$(s,a,s',p, p') \gets trace[i] $ \;
			\If(\tcp*[f]{if yes, store the step}){$p = n$}{
				$reachedFinal \gets \top $,
				$reachedStep \gets i$\;
				$break$\;
			}
		}
		\BlankLine
		\For(\tcp*[f]{backward update for each step}){$ i = length(trace) \cdots 1 $}{
			$(s,a,s',p, p') \gets trace[i] $ \;
			$t \gets V_p(s,a)+1$,
			$V_p(s,a) \gets t$\;
			$explR \gets \frac{1}{t}$ \tcp*[f]{visits-based bonus}
			\;
			\If(\tcp*[f]{same predicate, update within a single Q-Table}){$p = p' \lor p = n$}{
				\If{$ i < length(trace) $}{
					$Q_p(s,a) \gets (1-\alpha) \cdot Q_p(s,a) + \alpha \cdot max(explR, \gamma \cdot \max_{a'}Q_p(s',a'))$ \;
				}
				\Else{
					$Q_p(s,a) \gets (1-\alpha) \cdot Q_p(s,a) + \alpha \cdot max(explR, 0)$ \;
				}
			}
			\BlankLine
			\Else(\tcp*[f]{sequence transitioned to a different predicate}){
				\leIf(\tcp*[f]{predicate progress bonus}){$p' > p$}{$progR \gets 2$}{$progR \gets 0$}
				\If(\tcp*[f]{final predicate bonus}){$reachedFinal$}{
					$ d \gets reachedStep - i  - 1$ \;
					$finalR \gets \gamma^d \cdot 2 $ \;
				}
				$Q_p(s,a) \gets (1-\alpha) \cdot Q_p(s,a) + \alpha \cdot max(explR, \gamma \cdot (progR + finalR))$ \;
			}

			\
		}
	}
	\caption{\biasExpAlgo{} - $processEpisode$ method}
	\label{alg:bias_exp_algo_2}
\end{algorithm}

%% file: contents/transition_system.tex
\section{Environments from distributed systems}
\label{sec:transition_system}

\Final{As explained in Section~\ref{sec:sys_model}, to enable RL exploration, we model a distributed system as an MDP. 
This requires defining the three methods that characterize the environment: \textit{reset}, \textit{actions}, and \textit{step}. 
Before describing these methods, we first define the state of the environment and the set of actions.
}
We choose our actions to represent the network configurations.
The reason for our choice is due to the limitations of testing distributed system implementations---we only control the network between the processes. 
Specifically, we restrict the set of possible network configurations to partitions between processes. 
The state of the environment contains two components: 
an abstraction over local states of processes and the partition configuration. 
Each abstract local state is identified by a ``color'' that excludes process identifiers. 
An example state is $\left\{\left\{c_1\right\},\left\{c_1, c_2\right\}\right\}$, 
where $c_1, c_2$ are abstract local states and one of the proceses with abstract state $c_1$ is isolated from the rest.
\Final{Intuitively, the \textit{reset} method sets the state of the environment 
to $\left\{\left\{c_0, c_0, c_0\right\}\right\}$ and restarts the node processes. 
The \textit{actions} method returns the possible network partitions from a given state. 
Finally, the \textit{step} method delivers messages and updates the state abstractions with new colors.}
In this section, we formalize \Final{the semantics of a state, action and the \textit{step} method. 
Additionally, we} provide general modelling guidelines for using RL on distributed systems.

We will use Raft as a running example of a distributed protocol. 
Raft strives to achieve consensus in a set of $n$ processes while tolerating at most $f \leq \frac{n}{3}$ crash failures. 
Rounds in Raft are identified by a term number. 
At the beginning of a term, processes try to elect a \textit{Leader}. 
Processes vote Yes to a \textit{Candidate} unless they have already voted to a different Candidate. 
If a Candidate receives a majority of votes, it becomes the leader and informs all the processes of a successful election. 
The remaining processes now become \textit{Followers}. 
If a process does not hear from a leader within a specified time, 
it increments the term number and becomes a Candidate for the new term. 
The protocol dictates the local state components of each process: (1) the term number, (2) the state of the process $\left\{Candidate, Leader, Follower\right\}$, and 
(3) the process which it has voted for. 
In industrial implementations of Raft however, process states contain lot more information. 
The color abstraction we use in our model picks only specific components of the local state to be included in the abstraction.

\subsection{Defining the MDP}

The global state of a distributed system consists of two components: an abstraction over local states ($ls_i \in LS$ for process $i$) and the network configuration. 
The actions correspond to changing the network configuration resulting in a new state. 
Specifically, we restrict ourselves to network configurations where processes are partitioned. 
Only processes in the same partition can exchange messages. 
An abstract local state in $s$ is represented by a color ($c \in \colors$) defined by a coloring function $\mathbb{C}: LS \rightarrow \colors$. The colors do not contain the identities of the processes. The network configuration in $s$ is the partition configuration stored as a multi set of multi sets of colors.

For example, consider the state $\left\{\left\{c_1\right\}, \left\{c_1, c_2\right\}\right\}$. 
In this state, two processes with abstract local states $c_1$ and $c_2$ can communicate with each other while the third process (also with abstract state $c_1$) is isolated from them.
The corresponding set of actions \Final{(returned by the \textit{actions} method)} are all possible partition configurations. 
For example $\left\{\left\{c_1, c_1\right\}, \left\{c_2\right\}\right\}, \left\{\left\{c_1, c_1, c_2\right\}\right\}, \cdots$. 
By picking an action \Final{(invoking the \textit{step} method)}, the exploration algorithm determines the new partition configuration in the resulting state, e.g., 
with action $\left\{\left\{c_1, c_1\right\}, \left\{c_2\right\}\right\}$, the two processes with current abstract state $c_1$ are grouped in the same partition 
and the third process $c_2$ is isolated from them.

However in the resulting state, the colors of the processes may be different. 
Based on the action, we deliver messages allowed by the partition configuration and drop the remaining messages. 
The new messages received (or not), determines the new colors of the processes. 
More formally, we define an action as follows. 
Given a state $s$ with cardinalities of colors $C_s: \colors \rightarrow \mathbb{N}$, the actions enabled in $s$ are all possible multi-sets of multi-sets of colors where the cardinality of each color $c\in\colors$ is exactly $C_s(c)$. 
Then, a transition from $s$ with some action can lead to a state $s'$ where the cardinality of colors $C_{s'}$ is arbitrarily different from $C_s$.

\begin{figure*}
    \centering
    \begin{subfigure}{0.33\textwidth}
        \centering
        \includegraphics[width=\linewidth]{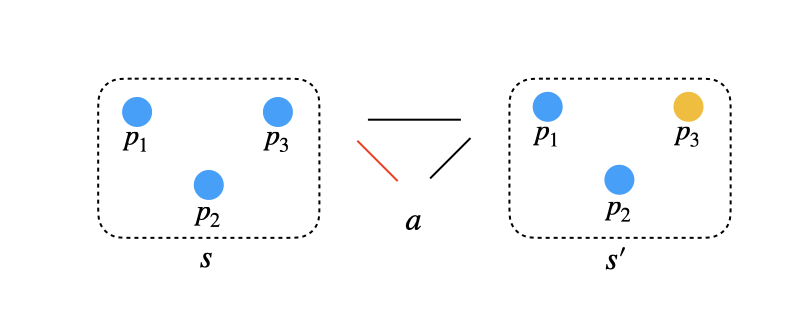}
        \caption{}
        \label{fig:dist_state_evol_a}
    \end{subfigure}%
    \begin{subfigure}{0.33\textwidth}
        \centering
        \includegraphics[width=\linewidth]{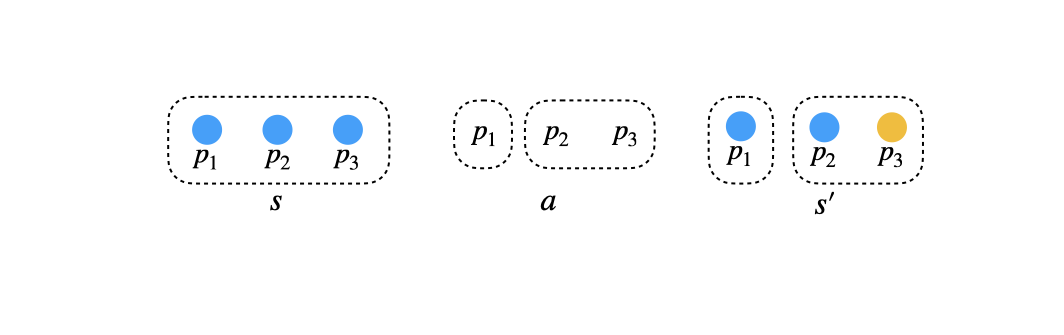}
        \caption{}
        \label{fig:dist_state_evol_b}
    \end{subfigure}%
    \begin{subfigure}{0.34\textwidth}
        \centering
        \includegraphics[width=\linewidth]{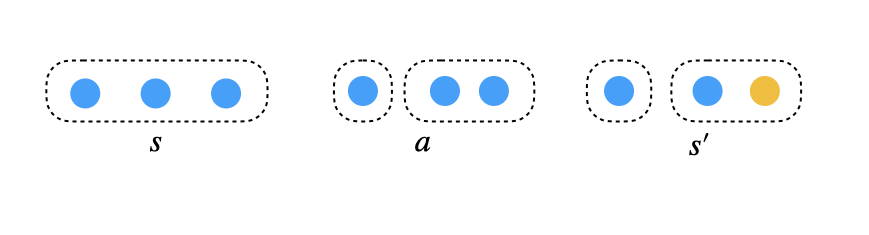}
        \caption{}
        \label{fig:dist_state_evol_c}
    \end{subfigure}%
    \caption{Evolution of the transition system of a distributed system. (a) Fine state space. State is a map of local states and action is the set of messages to deliver and drop. (b) Coarse state space without symmetry reduction. (c) Coarse state space with symmetry reduction}
    \label{fig:dist_state_evol}
\end{figure*}

Let us imagine a few simpler transition system models and use the drawbacks to motivate our final transition system. The first model is illustrated in Figure~\ref{fig:dist_state_evol_a}. The global state (3 blue dots with process identities) contains just a map of the complete local state and the network configuration is stored as a set of active network links.
The actions pick the next set of active network links (identified by non red arrows between processes). The resulting state (with process $p_3$ in a different state) is obtained by exchanging messages only along the active links. The granularity of actions is very fine with this model and therefore RL needs more steps and episodes to learn to reach new states. For example, reaching a new state with a quorum of messages requires learning to keep enough links active. The second model (Figure~\ref{fig:dist_state_evol_b}) introduces higher order actions where only partitions between processes are the valid network configurations. Each action corresponds to selecting a partition between processes (depicted using boxes around processes). The resulting state is obtained by exchanging only those messages within the same partition defined by the action. Such a model considers process identities when comparing two different states. Distributed protocols however define transitions only based on the state of the process and not the identities. Therefore, the drawback of the second model is to count redundant states where processes's local states are swapped around. We overcome the drawback by ignoring process identities and only considering an abstract local state (as colors) --- leading to our original definition of a state (Figure~\ref{fig:dist_state_evol_c}). An action re-defines the partition of colors in the pre-state (isolating the left-most process from the rest), but the processes receiving messages allowed by the new partition may result in colors changing in the post-state (e.g., the right-most process).

\subsection{General modelling guidelines}
\label{sec:modelling_guidelines}

RL exploration based algorithms perform best when the transitions in the state space are predictable \Final{(i.e. it behaves as an MDP)}. Specifically, the non-determinism in the transitions follow an underlying probability distribution that can be learnt. Therefore, we allow RL to perform better by reducing the degree of non-determinism in the transitions. The principle of making systems ``more'' deterministic however is also common with other testing approaches. With a more deterministic system, any bugs found can be easily replayed. 

In distributed protocols, time is the major contributor for non-determinism. The next local state of a process depends on an internal clock that is not modelled as part of the state. To remove this non-determinism, we fix the duration of time that passes between two transition steps in our transition system. 

Another source of non-determinism is the color abstraction. If the color abstraction excludes crucial state information, the transitions become unpredictable and therefore decrease the performance of RL based exploration algorithms. Consider the Raft protocol. If we do not include the role of each process in the color, RL will not be able to differentiate when a leader is elected or not. Furthermore, it will not able to learn to make progress and commit entries without observing that a leader is necessary to do so.
On the contrary, if we include too much information in the color abstraction, we will explode the state space that needs to be covered and also achieve sub-optimal outcomes. With redundant information in the color abstraction, we will reverse the effect of the symmetry reduction. For example, we should not include the process identifier in the color abstraction. 

Another general principle that we rely on is that the state space should be bounded. In other words, the set of possible colors should be bounded. Consider the raft protocol, each process consists of a term number as part of its local state. By including the term number in the color abstraction, we enable an infinite set of colors. Furthermore, we introduce redundancy. Consider two global states where the only difference is an offset in the term number of each local states. Only one of them is \textit{interesting} as the set of possible next states are the same modulo the offset. Therefore, the color abstraction should bound the term number of each local state. In general, it is important to balance the trade off when defining the color abstraction. Too much information explodes the state space and too little information introduces unpredictability.

Protocol implementations contain more information as part of the local state than what is defined in a model of the protocol. The additional information corresponds to optimizations introduced in the implementation. For example, most distributed protocol implementations introduce snapshots. When defining the color abstraction, the developer has to ensure that relevant parts of the additional state are also captured. As we will describe in Section~\ref{sec:hierarchies}, capturing the additional information enables the developer to bias exploration and test the parts of the code related to the optimizations.

\subsection{Environment parameters}

Our final transition system used in the implementation of our approach contains a few more optimizations.

\begin{enumerate}
    \item  We introduce failure actions to crash and start processes. The color abstraction contains an additional parameter to depict if the process is crashed or active.
    \item We introduce a finite number of actions where RL can inject new requests to the system. With new requests, we are able to explore more states that wouldn't be possible otherwise. For example with raft, the commit index increases only when new requests are committed.
    \item We introduce a parameter \textit{ticks} to control the time duration between two states. The number of ticks is tied to the timeout parameters of the protocols. If too much time passes between two states, then processes always timeout and if too little time passes then processes never timeout. The parameter allows the test developer to control the tradeoff and explore more states.
    \item We introduce a bounded counter \textit{SameState} in the global state representation, which increments, up to its bound, if the state (colors configuration) did not change after keeping the same partition. This incentivizes RL to explore the same state up to the counter limit, considering them different states. Setting a short time duration between states enables fine-grained interleavings of different partitions, potentially leading to new states. On the other hand, the protocol might require multiple steps in the same partitions configuration to progress. The counter allows to set a short time duration while enabling RL to explicitly explore scenarios of partitions stability.
\end{enumerate}

In total, to test an implementation, the developer has to specify the following parameters. We will refer to these parameters in Section~\ref{sec:eval} when we list the concrete values used for the different benchmarks.

\begin{enumerate}
    \item \textbf{The color abstraction}. The components of the state that define the color of each process.
    \item \textbf{Number of processes}. The number of processes in the system.
    \item \textbf{Ticks}. The \textit{ticks} parameter explained above.
    \item \textbf{Max Same State}. The maximum value for the \textit{SameState} counter.
    \item \textbf{Max Crash Actions}. The maximum number of crash actions allowed in an episode.
    \item \textbf{Max Concurrent Crashes}. The maximum number of processes that can be crashed at the same time.
\end{enumerate}

%% file: contents/hierarchies.tex
\section{Predicate sequences}
\label{sec:hierarchies}

Unlike with the cube world, the distributed system state space is significantly larger and harder to visualize. Pure exploration is insufficient to cover all possible states. Therefore, the need to bias exploration using \biasExpAlgo{} is all the more relevant. To bias exploration in distributed systems, we will use state predicates to define both the target state space and the intermediate rewards. In the Raft protocol, an example predicate would be - `there exists a leader'. A predicate captures a set of executions scenarios. If we strengthen the predicate - `there exists a leader in term 3' - we will constrain the set of admitted scenarios. Biasing exploration towards specific scenarios achieves two purposes. The developer gains confidence in the code when there are no bugs found and if there are bugs found, biased exploration will reliably reproduce the bugs. We rely on the developers understanding of the protocol to provide target predicates. In this section, we list example predicates for Raft and provide general guidelines to derive predicates for distributed protocols.

\subsection{Deriving target predicates}
As clear from the example, our main source of target predicates is the abstract protocol specification. The developer's understanding of the protocol specification is sufficient to construct scenarios and use them to bias exploration. However, a developer will have to instantiate the predicates by accessing the data structures used in the implementation.

\input{tables/predicates_raft.tex}

In general, a scenario of a distributed protocol contains segments of two kinds. One where processes are in sync and one where processes are out of sync. Using this insight, we derive three classes of predicates - ones which describe processes in sync, ones which describe processes out of sync, a combination of the two. We will refer to the examples of Raft protocol listed in Table~\ref{tab:predicates_raft}. Progress occurs when processes are in sync - by committing entries (CommittedEntries(2)), having a stable leader (LeaderInTerm(2)). However, progress stalls when processes are out of sync - processes in different terms (ProcessesInTerm(1,2) and ProcessesInTerm(1,4)), difference in logs (LogDiff(2)). The interesting scenarios are the ones where we combine the two. For example, a leader in a higher term (ProcessInRoleTerm(leader,3) and ProcessesInTerm(1,1)), difference in committed entries (LogCommitDiff(2)). Note that by \textit{sync} or \textit{out of sync}, we are referring to the abstract notion captured in the states of each process and not the concrete network state. 

Note that our list of predicates is not exhaustive and the developer is unconstrained while listing target predicates. Apart from the protocol, the developer can derive scenarios based on implementation specific optimizations. As mentioned in Section~\ref{sec:modelling_guidelines}, the implementation introduces some specific optimizations to the protocol such as snapshots and recovery. An example scenario would be to force a snapshot - `process with snapshot index 2'. Note that these predicates can be combined with those derived from the protocol.

\subsection{Specifying intermediate predicates}

When biasing exploration using a target predicate, the ability of \biasExpAlgo{} to effectively bias depends on how many scenarios the target predicate captures. As we will show in our evaluation, when the predicates are easy to satisfy \biasExpAlgo{} out performs other approaches with just one target predicate. However, a more constrained predicate is harder to bias towards. Therefore, \biasExpAlgo{} accepts additional intermediate predicates as waypoints to improve the effectiveness of biasing exploration. The intermediate predicates are used to split the task of reaching the target predicate to provide intermediate rewards to RL. For those target predicates with intermediate waypoints, we will empirically show that providing more intermediate predicates improves the effectiveness of biasing exploration. The question now arises on how to derive the intermediate predicates given a set of target predicates.

In Table~\ref{tab:predicates_raft}, we also list candidates for intermediate predicates. In general, predicates that are true on every execution path towards the target space are good candidates. For example, consider ProcessesInTerm(1, 3) where we require a process in term 3. To achieve the target, we need ProcessesInTerm(1, 2) to be true first which serves as an intermediate predicate. 

%% file: tables/predicates_raft.tex
\begin{table}[]
	\centering
	\small
	\begin{tabular}{|c|c|c|}
		\toprule
		\textbf{Hierarchy Name} & \textbf{Description} & \textbf{Intermediate Predicates} \\ 
		\midrule
		AllCommitted(x)          & \begin{tabular}[c]{@{}c@{}}At least x committed entries \\ in the log of all processes\end{tabular}          & AllCommitted(x - i)                                                                    \\
		\hline
		ProcessesInTerm(n, t) & \begin{tabular}[c]{@{}c@{}}At least n processes are \\ in term t\end{tabular} & ProcessesInTerm(n, t-1) \\
		\hline
		CommittedEntriesInTerm(x, t) & \begin{tabular}[c]{@{}c@{}}At least x committed entries \\ in the logof a process in term t\end{tabular} & \begin{tabular}[c]{@{}c@{}}CommittedEntriesInTerm(x - i, t)\\ LeaderInTerm(t)\end{tabular} \\
		\hline
		LeaderInTerm(t)              & \begin{tabular}[c]{@{}c@{}}A process is in state 'leader' \\ in term t\end{tabular}                      & ProcessesInTerm(n, t) \\
		\hline
		LogDiff(x)             & \begin{tabular}[c]{@{}c@{}}A gap of x entries between any \\ two processes logs\end{tabular}    & LogDiff(x - i) \\ 
		\hline
		LogCommitDiff(x)             & \begin{tabular}[c]{@{}c@{}}A gap of x entries between any \\ two processes committed logs\end{tabular}    & \begin{tabular}[c]{@{}c@{}}LogDiff(x - i)\\ LogCommitDiff(x - i)\end{tabular}              \\
		\hline
		ProcessInRole(r)      & Any process in role r              & -                                                               \\ 
		\hline
		ProcessInRoleTerm(r,t)      & \begin{tabular}[c]{@{}c@{}}A process in role r \\ and in term t\end{tabular}            & ProcessesInTerm(1, t)                                                               \\ \bottomrule
	\end{tabular}
	\caption{Generic predicates for Raft}
	\label{tab:predicates_raft}
\end{table}

%% file: contents/eval.tex
\section{Evaluation}
\label{sec:eval}

We evaluate the algorithms \pureExpAlgo{} and \biasExpAlgo{} on three benchmarks: RSL, Etcd, and RedisRaft. 
Our Benchmarks are (1) RSL - a re-implementation of Azure RSL~\footnote{https://github.com/Azure/RSL}. 
The algorithm is a variant of Paxos and powers distributed services in the Azure cloud. 
(2) Etcd~\footnote{https://github.com/etcd-io/raft}, an implementation of the Raft protocol that powers a popular distributed key value store. 
(3) RedisRaft~\footnote{https://github.com/RedisLabs/redisraft}, a  distributed version of the popular in-memory key value store that uses the Raft protocol. 
We re-implement RSL to enable instrumentation and testing on a common Linux platform. The original implementation is built to run on Windows systems.

We implement\footnote{https://anonymous.4open.science/r/rsl\_testing-C378/README.md} our RL algorithms in the Go programming language. 
In addition to the algorithms, our codebase consists of thin shims around the implementations to enable testing. 
The shim allows RL to read the state and execute the actions chosen at each step. 
\Final{Specifically, the shim captures the messages in flight by wrapping the transport interfaces and also interacts with the existing API to read node state. Due to its generic nature, the shim can be reused to implement other state exploration algorithms. Quantitatively, the shim instrumentation for RedisRaft benchmarks consists of 700 LOC (10k LOC total codebase) and for Etcd 500 LOC (7k LOC total codebase).} 

We compare against two baseline approaches, Random and \negRL{}~\cite{DBLP:journals/pacmpl/MukherjeeDBL20}, in all our experiments. 
Random implements pure random exploration, which picks the next actions uniformly at random, 
and \negRL{} is RL with a negative reward at each step. 
We implement the \negRL{} policy in our system. 
It performs Q-values updates with negative rewards at each step and uses the \textit{softmax} function to pick actions. 
The original description of the reward is a constant value of -1. 
However, the authors of \negRL{} describe an alternative version \negRLVisits{} which proves to be empirically better. 
We compare against \negRLVisits{} where the reward for a step that reaches a state $s$ is the negation of the number of visits to state $s$. 

With our evaluation, we aim to answer the following research questions. 
\begin{enumerate}
    \item[\textbf{RQ1}] Can we achieve better coverage than existing approaches with \pureExpAlgo{}? 
    \item[\textbf{RQ2}] Can we bias exploration towards a target state space with \biasExpAlgo{}?
    \item[\textbf{RQ3}] Does RL based exploration approaches help uncover bugs? 
\end{enumerate}

For \textbf{RQ1}, we find that \negRLVisits{} achieves better coverage than RSL in RedisRaft and Etcd benchmarks. However, \pureExpAlgo{} performs better than \negRLVisits{} in the RSL benchmark. The answer to \textbf{RQ2} is Yes, \biasExpAlgo{} explores more unique states in the target space than pure exploration approaches for 20 out of 26 different target predicates. Furthermore, when the target states are harder to reach, we show that we can improve the coverage by adding more intermediate predicates. For \textbf{RQ3}, we find 13 bugs with \biasExpAlgo{} as opposed to 11 with Random, 10 with \pureExpAlgo{} and 7 with \negRLVisits{}. Additionally, for 11 of these bugs, we are able to replicate the bugs with higher average frequency using \biasExpAlgo{} than other approaches. 

In the rest of the section, we first describe the coverage metric and the test harness parameters. Then, we present our evaluation for the three research questions.

\subsection{Test setup}
In our experimental results, we present comparison between different approaches using a coverage metric. The coverage metric measures the number of unique abstract states observed in each of the benchmarks. Specifically, the abstract state we measure is a multi-set of colors $s \subseteq (\colors \times \mathbb{N})^n$ where colors are abstract local states of each process.  
While the concrete abstraction of local states differs between the different benchmarks, we follow the same principles when abstracting. Namely, the color abstraction includes
\begin{enumerate}
	\item A round number (term, ballot, round, etc.)
	\item The role (leader, proposer)
	\item The log of requests 
	\item A commit index (commit, number of decided entries)
	\item Current vote or leader
\end{enumerate}

To run experiments, we need to tune two sets of parameters. One is related to the environment and the other related to the exploration algorithm. 
Here we list the concrete values used for both sets. The values for the environment parameters described in Section~\ref{sec:transition_system} are as follows,
\begin{enumerate}
	\item Number of nodes in the system - 3
	\item Ticks between steps, controls the time duration passed when executing an action on the system - 4 units
	\item \textit{SameState} counter limit - 5 
	\item Maximum number of crash actions in an episode - 3
	\item Maximum number of nodes to be crashed at the same time - 1
\end{enumerate}

The parameter values for the RL based polices described in Section~\ref{sec:overview} are as follows (the values for \negRLVisits{} are chosen according to the recommendations of the authors),
\begin{enumerate}
	\item The learning rate $\alpha$ is $0.2$ for both \pureExpAlgo{} and \biasExpAlgo{}, and $0.3$ for \negRLVisits{}.
	\item The discount factor $\gamma$ is $0.95$ for both \pureExpAlgo{} and \biasExpAlgo{}, and $0.7$ for \negRLVisits{}.
	\item The $\epsilon$-greedy values for both \pureExpAlgo{} and \biasExpAlgo{} is $0.05$
\end{enumerate}

Our experiments are run for 10000 episodes or 8h whichever occurs first. Each episode has an horizon of 25 in Etcd and RSL (max 250k total timesteps) and 50 in RedisRaft (max 500k total time steps). Horizon is set such that we are able to reach all possible target states and the difference in horizon is due to the different granularity of steps supported by the implementation. However, some episodes might terminate without running for the entire horizon due to a failure in the underlying system. Therefore, when comparing different approaches, we plot the coverage metric against the number of time steps passed instead of number of episodes. We conduct at least 10 trials for each benchmark and report the average numbers. To claim statistical significance in higher coverage, we perform the standard 
Mann-Whitney U test with a threshold of $>0.05$. 

\subsection{\textbf{RQ1}: Can we achieve better coverage with \pureExpAlgo{}?}

\begin{figure}
    \centering
	\begin{subfigure}{0.5\textwidth}
		\centering
		\includegraphics[width=\linewidth]{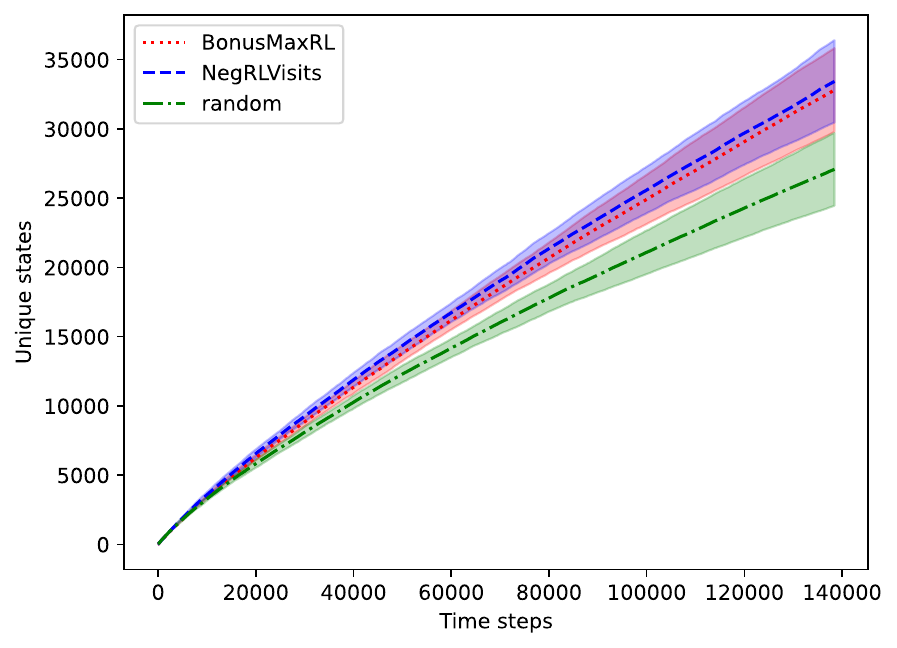}
		\caption{RedisRaft}
	\end{subfigure}%
	\begin{subfigure}{0.5\textwidth}
		\centering
		\includegraphics[width=\linewidth]{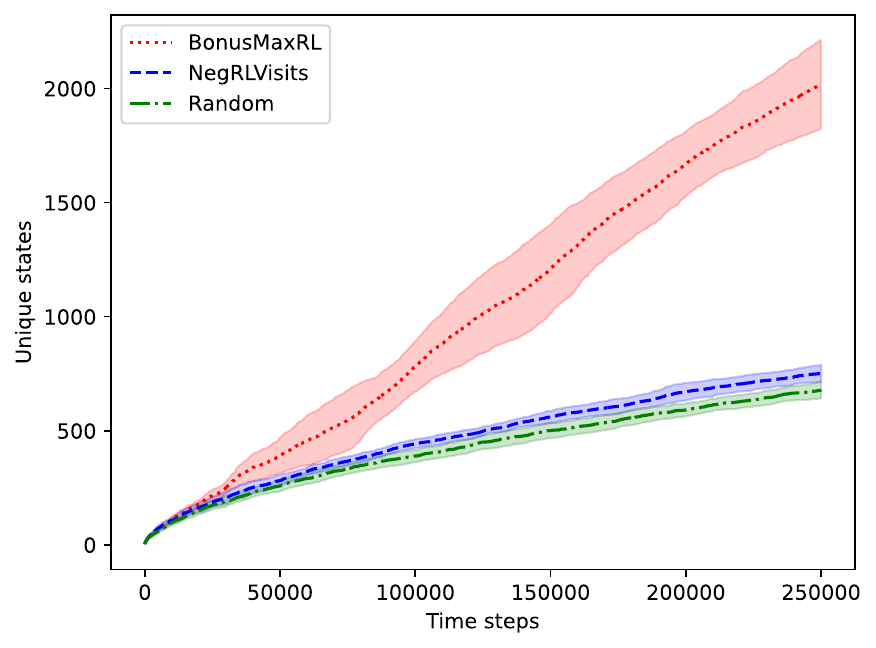}
		\caption{RSL}
	\end{subfigure}%
	\caption{The pure coverage comparison between \negRLVisits{}, \pureExpAlgo{} and Random for the different benchmarks. Each plot contains the average coverage vs time steps}				
    \label{fig:pure_coverage}
\end{figure}

\input{tables/coverage.tex}

We observe that RL based approaches achieve significantly better coverage than random exploration. Between the two RL approaches, \pureExpAlgo{} achieves better coverage than \negRLVisits{} approaches in the RSL benchmark and \negRLVisits{} has better coverage in the remaining 2.  We present in Figure~\ref{fig:pure_coverage} the coverage of the different exploration algorithms for RedisRaft and RSL benchmarks. The results of the Etcd benchmark are similar to RedisRaft. We believe that the high negative reward in \negRLVisits{} provides a very strong incentive for pure exploration and hence the better overall coverage. Table~\ref{tab:pure_coverage} contains the final average coverage numbers of the three approaches for the different benchmarks. Overall, \pureExpAlgo{} covers 21\%, 15.7\%, 197\% more states than Random in the RedisRaft, Etcd and RSL benchmarks respectively and 169\% more than \negRLVisits{} in the RSL benchmark. \negRLVisits{} covers 12.1\% more states than \pureExpAlgo{} in Etcd benchmark. Both \negRLVisits{} and \pureExpAlgo{} achieve similar coverage in RedisRaft benchmark.

While \negRLVisits{} is able to achieve higher coverage than \pureExpAlgo{} (except in the RSL benchmark), it is unclear if the reward mechanism can be adopted to bias exploration. Our reward mechanism to bias relies on providing a constant positive reward. \negRLVisits{} does not limit the $Q$-values within a fixed range, they will get more and more negative with the increasing number of visits. In this scenario, it would be difficult to control the tradeoff between reward exploitation and further exploration. A low constant reward value could be unable to balance the negative rewards and hence fail to learn, while a high value might immediately drive the policy to converge to that path, giving up further exploration and optimization. Therefore, \biasExpAlgo{} is based on the principles used in \pureExpAlgo{}.

\Final{Recent work, \textsc{Mallory}~\cite{DBLP:conf/ccs/MengPRS23}, employs RL to test distributed systems. It relies on the reward augmentation of \negRLVisits{} 
to select the set of failures to inject and improve overall coverage. 
We omit an empirical comparison to \textsc{Mallory} due to unavailability of the annotated source code needed by \textsc{Mallory} to measure coverage. 
The annotations identify ``interesting'' events in the execution that \textsc{Mallory} uses to mark coverage.
}

\Final{
It will be interesting to compare the tradeoffs between \negRLVisits{} and \pureExpAlgo{} in the context of \textsc{Mallory}.
It may be more interesting to add waypoints to \textsc{Mallory}'s algorithm.
As we demonstrate below, generic exploration algorithms fail to explore specific parts of the state space and we need biased exploration using \biasExpAlgo{} to cover them.
}

\subsection{RQ2: Can we bias exploration towards a target state space with \biasExpAlgo{}?}

\input{tables/target_redis.tex}

We answer RQ2 positively.
Given a sequence of predicates, we are able to bias exploration to cover more states in the target coverage with \biasExpAlgo{}. 
As motivated in Section~\ref{sec:hierarchies}, we use a total of 26 target predicates for all the benchmarks together in order to bias exploration. 
Table~\ref{tab:targets_redis} describes the set of target predicates that we use to bias for the two Raft benchmarks and RSL benchmark. \Final{For Etcd and RedisRaft protocols, we only list those predicates not referred to in Table~\ref{tab:predicates_raft}}
As mentioned in Section~\ref{sec:hierarchies}, the predicates belong to classes which require processes to be in sync, out of sync or a combination of both.

\input{tables/results_redis.tex}

For each target predicate, we measure the number of unique states observed that appear in the episode after the final target predicate has been satisfied. Table~\ref{tab:results_redis} compares the target coverage of \biasExpAlgo{}, \pureExpAlgo{}, \negRLVisits{} and Random exploration. For 20 out of 26 predicates, we observe that the biased exploration \biasExpAlgo{} guided by the predicate achieves significantly\footnote{Using Mann-Whitney U test.} more states than all the unbiased approaches - \pureExpAlgo{}, \negRLVisits{} and Random. Furthermore, the difference in the biasing is more stark for those target predicates where the number of permitted scenarios is low. For example, consider the predicate LogCommitDiff3 in RedisRaft which requires that there are two processes whose commit indices differ by 3. \biasExpAlgo{} on average covers 97x, 94x, 147x more states than \pureExpAlgo{}, \negRL{} and Random respectively. For the predicates for which we did not achieve improved coverage, we can try to speculate about the reasons. A possible reason is that a predicate might be too easy to reach for the unbiased baselines. In these cases, the impact of learning a policy to reach the predicate is reduced, while, on the other hand, it can reduce the variability of the explored executions. This could be the case for a predicate as OneInTerm3, which can easily happen during any execution. Other predicates might be simply too hard to reach. In such cases, specific knowledge of the protocol implementation details could help with designing better intermediate predicates.

\paragraph{Sensitivity to intermediate rewards}
\input{tables/intermediate_preds}

\begin{figure}
	\centering
	\begin{subfigure}{0.5\textwidth}
		\includegraphics[width=\linewidth]{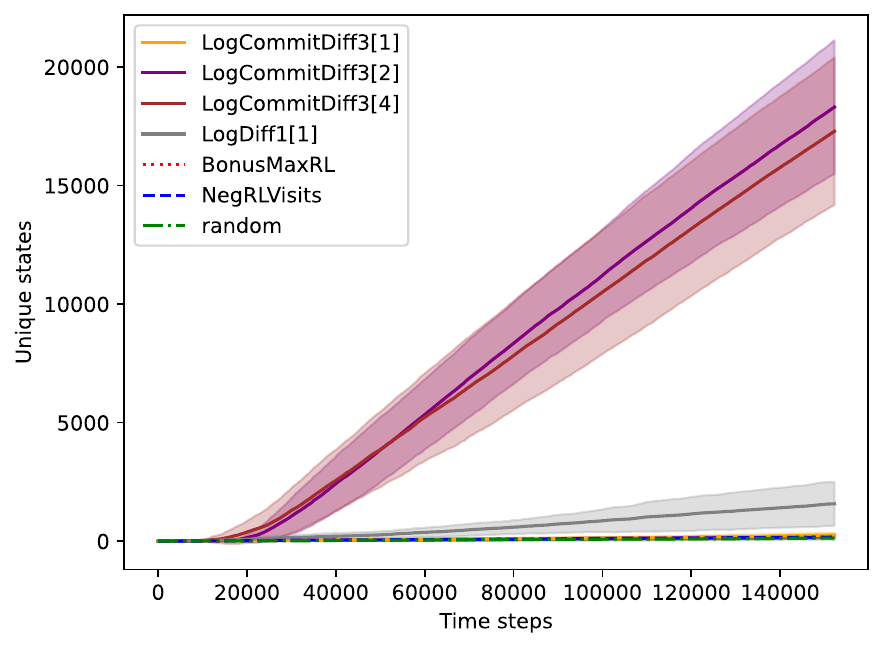}
		\caption{LogCommitDiff(3)}
		\label{fig:logcommitdiff3_1}
	\end{subfigure}%
	\begin{subfigure}{0.5\textwidth}
		\includegraphics[width=\linewidth]{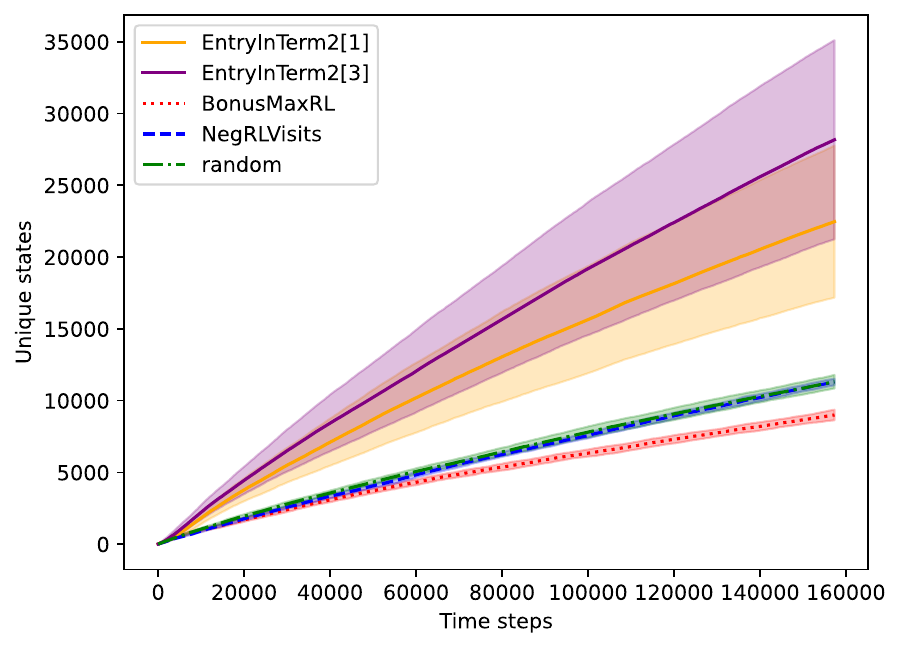}
		\caption{EntryInTerm(2)}
		\label{fig:entryinterm2}
	\end{subfigure}%
	\caption{Different versions of predicate sequences for the same target coverage. For the predicate sequences, the legend specifies the target predicate and, between square brackets, the number of predicates in the sequence excluding the first one (true predicate).}
	\label{fig:intermediate_comp}
\end{figure}

In Section~\ref{sec:hierarchies}, we describe a methodology to list intermediate predicates to bias the exploration along with example predicates for Raft (Table~\ref{tab:predicates_raft}). The intermediate predicates help \biasExpAlgo{} to reach the target space faster and improve the accuracy especially when the target space described by a predicate is hard to reach.  We list in Table~\ref{tab:intermediate_preds} the target coverage with increasing number of intermediate predicates for three target predicates. The result is also visualized in Figure~\ref{fig:intermediate_comp} for two of the target predicates. Let us consider the example of LogCommitDiff(3), where we use LogDiff(1), LogDiff(2), and LogDiff(3) as intermediate predicates. We observe that the biasing without intermediate predicates fails to improve coverage over the unbiased baselines in the given time budget. On the other hand, when using respectively 1 (LogDiff(1)) and all the 3 intermediate predicates, the final coverage is significantly better, showing that intermediate predicates can play a key role in the biased exploration efficiency and performance. Interestingly, by providing LogDiff(1) as the target predicate, we still achieve improved coverage over the unbiased baselines, showing that even partial biasing can potentially improve a different target coverage. 

\subsection{RQ3: Does biased exploration help uncover bugs?}

\input{tables/redis_bugs_occurs.tex}

We evaluate the ability of different approaches to find bugs by measuring the number of bugs and the average occurrence of each bug. As shown in Table~\ref{tab:redis_bug_occurs}, we find that RL based approaches are better at reliably replicating bugs with biased exploration having the best outcomes. The table lists the average occurrence of each bug. Furthermore, biased exploration is able to uncover all the bugs while pure exploration is able to uncover only some.

\paragraph{RedisRaft}
\input{tables/redis_bug_descriptions.tex}
We are able to identify 3 new bugs and reproduce 7 known bugs in RedisRaft. Table~\ref{tab:redis_bug_desc} provides a short description of the bugs. We capture two classes of bugs for RedisRaft when testing with using \pureExpAlgo{} and \biasExpAlgo{}. First, bugs that violate a safety property of the protocol during an episode. Second, an unexpected failure in the implementation during the episode. For the first class of bugs, we are able to capture the trace and identify any issues. However analyzing an unexpected failure in the implementation code requires a deeper understanding of the implementation codebase. 

Using \biasExpAlgo{}, we are able to identify all bugs. However, pure exploration approaches fail to replicate one bug. Furthermore, for some bugs, the higher average occurrence using the predicate for biased exploration correlates with the bug description. For example, RaftAppendEntry occurs when biasing exploration to add an entry to the log, RaftBecomeFollower bug is more common when biasing exploration towards states where all processes transition to term 2.

\paragraph{Etcd}
Etcd is a robust and well tested implementation that has been used in production for many years. Despite the robustness, we replicate 1 known bug and find 1 new bug with etcd with our testing efforts. The new bug, that occurs due to incorrect log restoration, is uncovered by biased exploration as well as by random exploration with the same frequency. However, the known bug is uncovered only using biased exploration.

\paragraph{RSL}
We find one new bug (safety violation) in our RSL implementation where two processes decide on different values. The bug is caught by all approaches but replicated more frequently than other approaches using biased exploration.

%% file: tables/coverage.tex
\begin{table}
    \centering
    \begin{tabular}{|c|c|c|c|}
        \toprule
        Benchmark & Random & \pureExpAlgo{} & \negRL{} \\
        \midrule
        RedisRaft & 27081.9 $\pm$ 2627.31 & 32818.4 $\pm$ 3017.32 & 33433.2 $\pm$ 2972.34 \\
        Etcd & 19179.3 $\pm$ 106.26 & 22202.7 $\pm$ 137.75 & \textbf{24898.6 $\pm$ 97.40} \\
        RSL & 678.9 $\pm$ 36.43 & \textbf{2020.9 $\pm$ 190.67} & 751.2 $\pm$ 37.24 \\
        \midrule
    \end{tabular}
    \caption{Final average coverage values for the different benchmarks.}
    \label{tab:pure_coverage}
\end{table}

%% file: tables/target_redis.tex
\begin{table}[]
	\centering
	\small
	\begin{tabular}{|l|l|}
		\toprule
		\textbf{Target Predicate} & \textbf{Description}                                           \\ 
		\midrule
		\textbf{Raft (RedisRaft, Etcd) } & \\
		AllInTerm(2)              & All the processes simultaneously in term 2                     \\ 
		TermDiff(2)    & A difference of 2 terms between any two processes     \\ 
		CommitEntries(2)          & At least 2 committed entries in the log of a process           \\ 
		EntryInTerm(2)            & At least 1 committed entries in the log of a process in term 2 \\ 
		OneLeaderOneCandidate   & A process in state "leader" and another in state "candidate"   \\ 
		\hline
		\textbf{RSL} & \\
		AllBallot(3) & All processes reach ballot (round) 3 \\
		AnyBallot(3) & Any process in ballot 3 \\
		AnyDecided(3) & Any process decides 3 entries \\
		AnyDecree(2) & Any process reaches decree 2 \\
		BallotDiff(2) & A difference of 2 ballots between two active processes \\
		EntryBallot(2) & An entry in the log in ballot 2 \\
		PrimaryBallot(2) & A process becomes primary (leader) in ballot 2 \\
		DecidedDiff(3) & A difference of 3 decided log entries between any two processes \\
		\bottomrule
	\end{tabular}
	\caption{Target predicate descriptions used to bias exploration in Raft and RSL benchmarks}
	\label{tab:targets_redis}
\end{table}

%% file: tables/results_redis.tex
\begin{table}[]
	\centering
	\small
	\begin{tabular}{|l|c|c|c|c|c|}
		\toprule
		Benchmarks         & No.Pred & \biasExpAlgo{}          & BonusMaxRL       & NegRLVisits      & Random           \\
		\midrule
		\textbf{RedisRaft} & & & & & \\
		OneInTerm(3)                          & 1   &  15934 $\pm$ 3040  & 16640 $\pm$ 2033  & 17214 $\pm$ 2103  & 14370 $\pm$ 1810  \\
		AllInTerm(2)                          & 1   & \textbf{23695 $\pm$ 3810} & 7660 $\pm$ 945 & 8437 $\pm$ 876 & 6401 $\pm$ 794 \\
		TermDiff(2)                           & 1   & 23478 $\pm$ 2917 & 22214 $\pm$ 1973 & 23323 $\pm$ 2149 & 20112 $\pm$ 2007 \\
		CommitEntries(2)                      & 1   & \textbf{24656 $\pm$ 3995} & 4894 $\pm$ 552   & 5325 $\pm$ 741  & 2835 $\pm$ 530   \\
		EntryInTerm(2)                        & 3   &  \textbf{22758 $\pm$ 5457}  & 8267 $\pm$ 272  & 10114 $\pm$ 418  & 9812 $\pm$ 662  \\
		LeaderInTerm(2)                       & 1   & \textbf{22533 $\pm$ 5418} & 8971 $\pm$ 299  & 10834 $\pm$ 410  & 10361 $\pm$ 684  \\
		LogDiff(1)                            & 1   & \textbf{30779 $\pm$ 3365} & 5755 $\pm$ 713  & 5958 $\pm$ 825  & 3332 $\pm$ 606   \\
		LogCommitDiff(3)                      & 2   & \textbf{14960 $\pm$ 4100} & 154 $\pm$ 63     & 158 $\pm$ 52     & 102 $\pm$ 41     \\
		OneLeaderOneCandidate               & 3   & 1301 $\pm$ 1360  & 482 $\pm$ 120    & 471 $\pm$ 158    & 356 $\pm$ 93    \\
		\hline
		\textbf{Etcd} & & & & & \\
		LogCommitGap(3) & 4 & \textbf{13336 $\pm$ 664} & 5692 $\pm$ 132 & 6411 $\pm$ 112 & 4717 $\pm$ 92 \\
		OneInTerm(4) & 3 & \textbf{33011 $\pm$ 826} & 29545 $\pm$ 258 & 29739 $\pm$ 167 & 25309 $\pm$ 200 \\
		MinCommit(2) & 3 & \textbf{29862 $\pm$ 885} & 25015 $\pm$ 193 & 24031 $\pm$ 110 & 21765 $\pm$ 195 \\
		TermDiff(2) & 1 & \textbf{15289 $\pm$ 885} & 4673 $\pm$ 162 & 7792 $\pm$ 66 & 4879 $\pm$ 142 \\
		LeaderInTerm(4) & 3 & \textbf{15727 $\pm$ 1157} & 10684 $\pm$ 142 & 11171 $\pm$ 215 & 9571 $\pm$ 101 \\
		AtLeastOneCommitInTerm(2) & 3 & \textbf{32709 $\pm$ 1025} & 28169 $\pm$ 262 & 25863 $\pm$ 176 & 23903 $\pm$ 314 \\
		OneLeaderOneCandidate & 3 & 35403 $\pm$ 958 & 36178 $\pm$ 142 & 32891 $\pm$ 208 & 29021 $\pm$ 307 \\
		LogGap(2) & 2 & \textbf{37832 $\pm$ 3485} & 31372 $\pm$ 314 & 33040 $\pm$ 115 & 27445 $\pm$ 238 \\
		AllInTerm(5) & 3 & \textbf{10346 $\pm$ 1121} & 8202 $\pm$ 65 & 7888 $\pm$ 122 & 6682 $\pm$ 94 \\
		\hline
		\textbf{RSL} & & & & & \\
		AnyBallot(3) & 1 & \textbf{1573 $\pm$ 174} & 837 $\pm$ 75 & 301 $\pm$ 34 & 264 $\pm$ 31 \\
		AllBallot(3) & 1 & \textbf{1021 $\pm$ 57} & 493 $\pm$ 92 & 102 $\pm$ 14 & 101 $\pm$ 16 \\
		EntryBallot(2) & 1 & 1016 $\pm$ 460 & \textbf{1954 $\pm$ 131} & 698 $\pm$ 27 & 658 $\pm$ 48 \\
		AnyDecree(2) & 1 & \textbf{1068 $\pm$ 119} & 663 $\pm$ 52 & 188 $\pm$ 19 & 155 $\pm$ 23 \\
		BallotDiff(2) & 1 & 19 $\pm$ 5 & 12 $\pm$ 6 & 2 $\pm$ 2 & 3 $\pm$ 3 \\
		AnyDecided(3) & 1 & \textbf{856 $\pm$ 93} & 492 $\pm$ 46 & 134 $\pm$ 18 & 110 $\pm$ 16 \\
		PrimaryInBallot(2) & 2 & \textbf{607 $\pm$ 66} & 467 $\pm$ 54 & 232 $\pm$ 22 & 196 $\pm$ 30 \\
		DecidedDiff(3) & 3 & \textbf{113.7 $\pm$ 46.8} & 22.7 $\pm$ 10.1 & 5.7 $\pm$ 3.1 &  2.9 $\pm$ 1.5 \\
		\bottomrule
	\end{tabular}
	\caption{Coverage results - the table shows the target coverage results in our benchmarks. Each row contains the target predicate, the number of predicates in the sequence used by \biasExpAlgo{} (excluding the first one), and, for each algorithm, the average number of unique explored states ($\pm$ Standard Deviation).}
	\label{tab:results_redis}
\end{table}

%% file: tables/intermediate_preds.tex
\begin{table}
    \centering
    \footnotesize
    \begin{tabular}{|c|c|p{6cm}|c|}
        \toprule
        Benchmark & Target predicate & Predicates Sequence & Target Coverage \\
        \midrule
        \multirow{4}{*}{RedisRaft} & \multirow{4}{*}{LogCommitDiff(3)} & LogCommitDiff(3) & 264 $\pm$ 99 \\
        & & LogDiff(1), LogCommitDiff(3) & \textbf{18309 $\pm$ 2820} \\
        & & LogDiff(1), LogDiff(2), LogDiff(3), LogCommitDiff(3) & \textbf{17287 $\pm$ 3102} \\
        & & LogDiff(1) & 1578 $\pm$ 931 \\
        \hline
        \multirow{2}{*}{RedisRaft} & \multirow{2}{*}{EntryInTerm(2)} & EntryInTerm(2) & 22468.1 $\pm$ 5295.2 \\
        & & OneInTerm(2), LeaderInTerm(2), EntryInTerm(2) & 28173.7 $\pm$ 6935.9 \\
        \hline
        \multirow{3}{*}{RSL} & \multirow{3}{*}{DecidedDiff(3)} & DecidedDiff(3) & 26.4 $\pm$ 13.2 \\
        & & DecidedDiff(2), DecidedDiff(3) &  \textbf{173.1 $\pm$ 70.1} \\
        & & DecidedDiff(1), DecidedDiff(2), DecidedDiff(3) & \textbf{113.7 $\pm$ 46.8} \\
        \bottomrule
    \end{tabular}
    \caption{Improvements with intermediate predicates. For each target predicate (one example from each benchmark) and the sequence used to bias, we report the average final target coverage ($\pm$ Standard Deviation). We use the same classes of predicates described in Table~\ref{tab:targets_redis}, eventually instantiated with different values.}
    \label{tab:intermediate_preds}
\end{table}

%% file: tables/redis_bugs_occurs.tex
\begin{table}
    \centering
    \small
    \begin{tabular}{|l|c|c|c|c|}
        \toprule
        Bug & \biasExpAlgo{} & \pureExpAlgo{} & \negRLVisits{} & Random \\
        \midrule
        \textbf{RedisRaft} & & & &\\
        RaftRestoreLog & 5704.6 (AllInTerm(2)) & \textbf{6499.8} & 5056.2 & 4897.5 \\
        HandleBeforeSleep & \textbf{23.4 (LeaderInTerm(2))} & 13.8 & 15.1 & 7.4 \\
        ConnIsConnected & 1.3 (LogDiff(1)) & 0.5 & - & \textbf{4.5} \\
        RaftAppendEntry & \textbf{9.8 (EntryInTerm(2))} & 0.7 & 2.8 & 5.7 \\
        RaftBecomeFollower & \textbf{2.2 (AllInTerm(2))} & 0.2 & 0.8 & 1.5 \\
        RaftApplyEntry & \textbf{0.8 (LogDiff(1))} & 0.1 & - & 0.4 \\
        RaftDeleteEntry & \textbf{0.1 (LeaderInTerm(2))} & - & - & - \\
        InconsistentLogs & \textbf{2.0 (AllInTerm(2))} & 0.4 & 0.2 & 0.2 \\
        ReducedLogs & \textbf{7.1 (EntryInTerm(2))} & 2.2 & 2.1 & 1.9 \\
        ModifiedLog & \textbf{0.3 (LogDiff(1))} & 0.1 & - & 0.1 \\
        \hline
        \textbf{Etcd} & & & & \\
        IncorrectLogRestore & \textbf{0.1 (OneLeaderOneCandidate)} & - & - & - \\
        NilSnapshotPanic & \textbf{0.3 (LogCommitGap(3))} & - & - & 0.2 \\
        \hline
        \textbf{RSL} & & & & \\
        InconsistentLogs & \textbf{167.2 (AnyBallot(3))} & 51.1 & 8.3 & 8.0 \\
        \bottomrule
    \end{tabular}
    \caption{Average occurrence of bug comparing biased exploration with different pure-exploration algorithms. The column for \biasExpAlgo{} reports the highest average occurrence of the bug and the corresponding target predicate used to bias.}
    \label{tab:redis_bug_occurs}
\end{table}

%% file: tables/redis_bug_descriptions.tex
\begin{table}
    \centering
    \small
    \begin{tabular}{|l|l|p{6cm}|}
        \toprule
        Bug & Category & Description \\
        \midrule
        RaftRestoreLog & Crash & Occurs when restoring log from file \\
        HandleBeforeSleep & Crash & When flushing log entries to file \\
        ConnIsConnected & Crash & When connecting to a node added to the cluster \\
        RaftAppendEntry & Crash & When adding a new entry to the log \\
        RaftBecomeFollower & Crash & When updating the state to follower upon receiving an append entries message \\
        RaftApplyEntry & Crash & When applying a committed entry onto the state machine \\
        RaftDeleteEntry & Crash & When removing an uncommitted entry from the log \\
        InconsistentLogs & Safety violation & Two committed logs differ in an entry \\
        ReducedLogs & Safety violation & A process loses a committed entry in the log \\
        ModifiedLog & Safety violation & A process changes a committed entry in the log \\
        \bottomrule
    \end{tabular}
    \caption{Bug descriptions for RedisRaft benchmark along with the category of bug. Crash bugs are unexpected failures in the process and Safety violation bugs are those where the trace violates safety properties}
    \label{tab:redis_bug_desc}
\end{table}

%% file: contents/related_work.tex
\section{Related Work}
\label{sec:related}

\subsubsection*{Testing Distributed Systems.}
Over the years there have been many works to test distributed programs. Systematic exploration techniques~\cite{DBLP:conf/nsdi/YangCWXLLYLZZ09,DBLP:conf/spin/SimsaBG11,DBLP:conf/osdi/LeesatapornwongsaHJLG14} have failed to scale to the large state space of protocol implementations. Model based testing approaches such as Mocket~\cite{DBLP:conf/eurosys/WangDGWW023} face a similar challenge enumerating all possible failure scenarios. Randomized exploration techniques such as Jepsen~\cite{jepsen} have proven to be promising in uncovering bugs. Jepsen introduces arbitrary failures by crashing nodes or introducing network partitions. Other random techniques either provide probabilistic guarantees~\cite{DBLP:journals/pacmpl/OzkanMNBW18}, leverage partial order reduction techniques to shrink the search space~\cite{DBLP:journals/pacmpl/OzkanMO19,DBLP:journals/pacmpl/DragoiEOMN20}, use standard fuzzing by mutating test inputs~\cite{DBLP:conf/uss/ChenGXSZLWL20,DBLP:journals/pacmpl/WinterBGGO23,DBLP:conf/icse/GaoDWFWZH23,DBLP:conf/icse/MeertenOP23}. 

There are  approaches to \emph{verify} an implementation or to generate code from a verified instance \cite{DBLP:conf/sosp/HawblitzelHKLPR15,DBLP:conf/pldi/WilcoxWPTWEA15,DBLP:conf/osdi/ChajedTT0KZ21}, making testing unnecessary.
So far such implementations are seldom used in production applications.
Moreover, bugs can still lurk in such code at the interfaces between verified and unverified components \cite{DBLP:conf/eurosys/FonsecaZWK17}---thus, bug finding techniques remain essential
even when parts of the system are proved correct.

\subsubsection*{Reinforcement Learning for Testing.}
Two closely related works apply $Q$-learning techniques to test distributed or concurrent programs. QL~\cite{DBLP:journals/pacmpl/MukherjeeDBL20} applies Reinforcement learning techniques to test concurrent message passing and shared memory programs. They introduce a novel reward mechanism that provides strong incentives for exploration. However it is unclear if the reward mechanism can be extended to bias exploration. 

We note that RL has been used in testing in orthogoal ways, e.g., 
in sequential fuzzing \cite{DBLP:conf/icse/ReddyLPS20}, in learning appropriate parameters \cite{DBLP:conf/uss/WangZZQKA21}, synthesizing valid inputs~\cite{DBLP:conf/sp/BottingerGS18,DBLP:conf/fates/VeanesRC06}, and inputs that induce failures in control systems~\cite{DBLP:conf/kbse/ZhangLSCHLLH21}

\subsubsection*{Reward-Free Exploration in Reinforcement Learning.}
This line of work  provides theoretically efficient algorithms to explore a given environment in absence of a reward function~\cite{DBLP:conf/icml/JinKSY20,DBLP:conf/nips/ZhangMS20}. They provide theoretical guarantees over the coverage of the state space, also using decaying reward augmentation. Unfortunately, the number of episodes required for coverage are (5th degree) polynomial functions in the size of the states and actions and thus unsuitable in our setting.
When run for a limited amount of time, they do not perform well in practice compared to our algorithms and our baselines.
\Final{Furthermore, it is unclear how these techniques can be extended to bias exploration while retaining the guarantees. In contrast, our approach extends pure exploration techniques with reward augmentation (using programmer provided semantic knowledge) that prove to be effective in biasing exploration.}

\subsubsection*{Splitting Goals into Subgoals.} 
Hierarchical RL relies on splitting an RL problem into subtasks, learn policies to solve each subtask, and then combine these policies to solve the original problem 
\cite{DBLP:journals/ai/SuttonPS99,DBLP:journals/jair/Dietterich00,DBLP:conf/nips/ParrR97}.
These methods build on the idea of defining higher level actions, with multiple timesteps duration, to solve the subtasks. 
They then learn some sort of global policy that chooses which of these actions to follow at each timestep. 
While we also leverage the idea of splitting the problem into smaller tasks, our subtasks are just waypoints towards the goal with fixed priorities. 
Our approach is simpler and does not allow to reuse or combine these sub-policies in a structured way.

\subsubsection*{Reward Machines}
Another way to leverage tasks decomposition and knowledge of the reward function structure is given with Reward Machines ~\cite{DBLP:journals/jair/IcarteKVM22}. They allow for complex reward structure specification, hierarchical learning approach, and efficient algorithms to speed up policy optimization. Unfortunately, we can't benefit from these advantages in our setting. The efficient algorithms mainly build on the idea of decoupling the reward function from the environment transitions, allowing to simulate the result of an observed transition as if it happened at a different stage of the reward function. In our setting, the rewards strictly depend on the system transitions and hence we can't decouple them.

\subsubsection*{Temporal Goals in RL}
Recent work in combining temporal logic goals and reinforcement learning has also explored the idea of intermediate goals~\cite{DBLP:conf/nips/JothimuruganBBA21,DBLP:conf/aaai/JiangB0STS21,DBLP:conf/birthday/AlurBBJ22,DBLP:conf/ijcai/0005T19}. 
For example, given a goal described by an automaton, a subgoal is to reach intermediate states of the automaton between the initial and accepting states. Similar to the problem with reward machines, simulating transitions based on the subgoals is infeasible when testing real world implementations with minimal instrumentation.

%% file: contents/conclusion.tex
\section{Conclusion}
\label{sec:conclusion}

We have presented two new algorithms, \pureExpAlgo{} and \biasExpAlgo{}, for reward augmentation in RL for testing distributed systems.
We introduce a new reward augmentation mechanism in \pureExpAlgo{} that performs better than random exploration in our benchmarks. 
When combined with developer-provided waypoints, our \biasExpAlgo{} algorithm is quite effective in exploring the state space of protocol implementations. 
We discuss on a methodology to derive target predicates based on an understanding of the protocol. 
In total for the 3 benchmarks, we are able to bias exploration towards 26 different target state spaces. 
Furthermore, we are able to find new bugs and replicate known ones with \biasExpAlgo{}, often with a higher frequency than baseline approaches.